\documentclass[11pt,a4paper,english]{article}
\pdfoutput=1

\usepackage{lmodern}
\renewcommand{\sfdefault}{lmss}

\usepackage{color}
\usepackage{verbatim}
\usepackage{mathtools}
\usepackage{enumitem}
\usepackage{amsmath}
\usepackage{amsthm}
\usepackage{amssymb}
\usepackage{graphicx}

\makeatletter
\newif\ifprstyle

\ifx\affiliation\undefined
  \prstylefalse
\else
  \prstyletrue
  \newif\ifnotoc
  \notoctrue
\fi

\ifprstyle
  \usepackage[
    colorlinks=true, pdfa=true,
    urlcolor=blue,   anchorcolor=blue,
    citecolor=blue,  filecolor=blue,
    linkcolor=blue,  menucolor=blue,
    pagecolor=blue,  linktocpage=true
  ]{hyperref}
\else
  \usepackage{jheppub}
  \newcommand{\email}[1]{\emailAdd{#1}}
\fi

\ifx\ifCrop\undefined\else

  \ifx\ifPresentation\undefined
     \usepackage[paperwidth=171.2mm,paperheight=261.2mm]{geometry}
  \else
     \usepackage[paperwidth=330mm,paperheight=233mm]{geometry}
  \fi

  \divide\@tempdima\baselineskip
  \@tempcnta=\@tempdima
  \ifx\ifPresentation\undefined
     \hypersetup{pdfstartview={Fit},pdfpagelayout={SinglePage}}
  \else
    \hypersetup{pdfstartview={Fit},pdfpagelayout={SinglePage}}

    \usepackage{everypage}
    \AddEverypageHook{\tikz[remember picture,overlay]{%
      \draw [gray!10] ($(current page.east)+(-130mm,+180mm)$)
         -- ($(current page.east)+(-130mm,-100mm)$);}}
  \fi
\fi

\hypersetup{
    urlcolor=black!40!blue,   anchorcolor=black!40!blue,
    citecolor=black!40!blue,  filecolor=black!40!blue,
    linkcolor=black!40!blue,  menucolor=black!40!blue,
    pagecolor=black!40!blue
}

\usepackage{bookmark}

\usepackage{amsmath,amsfonts,amssymb,bm,cancel}

\usepackage{color}
\usepackage{colortbl}
\ifprstyle
  
\else
  
\fi

\newcommand{\bSe}{\begin{subequations}}
\newcommand{\eSe}{\end{subequations}}

\newcommand{\bWe}{\begin{widetext}}
\newcommand{\eWe}{\end{widetext}}

\allowdisplaybreaks

\usepackage{tikz}
\usetikzlibrary{shapes,arrows}
\usetikzlibrary{calc}
\usetikzlibrary{positioning}
\usetikzlibrary{decorations.pathreplacing}
\usetikzlibrary{shapes.misc}

\usepackage[scr,scaled=1.1]{rsfso}
\usepackage{eucal}

\DeclareMathAlphabet{\mathsfit}{\encodingdefault}{\sfdefault}{m}{sl}
\SetMathAlphabet{\mathsfit}{bold}{\encodingdefault}{\sfdefault}{bx}{sl}

\usepackage{float}

\floatstyle{plaintop}
\newfloat{bbox}{thp}{lop}
\floatname{bbox}{Box}

\renewcommand\floatc@plain[2]{\setbox\@tempboxa\hbox{{\@fs@cfont #1.} #2}%
\ifdim\wd\@tempboxa>\hsize {\@fs@cfont #1.} #2\par
\else\hbox to\hsize{\hfil\box\@tempboxa\hfil}\fi}

\ifprstyle\else
\fi

\newcommand{\emitFrontMatter}{
  \ifprstyle
    \begin{abstract}\myAbstract\end{abstract}
  \else
    \ifnotoc
      \abstract{\\[-1.64ex]\hphantom{\hspace{0.28cm}}%
        \parbox{1\columnwidth-0.28cm}%
        {\renewcommand\baselinestretch{1.05}\small\hspace{18mm}\myAbstract}}
    \else
      \abstract{\myAbstract}
    \fi
  \fi

  \keywords{\myKeywords}

  \ifprstyle\else
    \makeatletter
      \def\@fpheader{\myReleaseInfo}
    \makeatother
    \ifnotoc
      \compress
      \renewcommand\afterLogoSpace{}
      \renewcommand\afterSubheaderSpace{}
      \renewcommand\afterProceedingsSpace{}
      \makeatletter
      \renewcommand\ps@titlepage{}
      \makeatother
      \toccontinuoustrue
    \else
      \renewcommand\afterTocSpace{\medskip}
      \addtocontents{toc}{\protect\setcounter{tocdepth}{2}}
    \fi
  \fi

  \maketitle

  \ifprstyle\else
    \ifnotoc
      \hrule\bigskip\bigskip
    \else
      \flushbottom
    \fi
  \fi
}

\usepackage{titlesec}

\newcommand{\emitAppendix}{
  \phantomsection
  \addcontentsline{toc}{section}{Appendices}
  \addtocontents{toc}{\protect\setcounter{tocdepth}{2}}

  \makeatletter
    \def\toclevel@section{1}
    \def\toclevel@subsection{2}
  \makeatother

  \addtocontents{toc}{\protect\makeatletter}
  \addtocontents{toc}{\string\let\string\l@chapter\string\l@section}
  \addtocontents{toc}{\string\let\string\l@section\string\l@subsection}
  \addtocontents{toc}{\protect\makeatother}

  \titleformat{\section}{\normalfont\large\bfseries}{Appendix~\thesection~~~}{0em}{}

  \let\oldSection\section
  \renewcommand\section{\bookmarksetupnext{level=2}\oldSection}

  \appendix
}

\makeatother

\usepackage{babel}
\begin{document}

\newif\ifColors

\newif\ifShowSigns 

%%%%%%%%%%%%%%%%%%%%%%%%%%%%%%%%%%%%%%%%%%%%%%%%%%%%%%%%%%%%%%%%%%%%%%%%%%%%%
% Avoid the redefinitions of the commands if already defined
\ifx \ii \undefined

% Adjusted color flavors
\definecolor{red}{rgb}{1,0,0.1}
\definecolor{green}{rgb}{0.0,0.6,0}
\definecolor{blue}{rgb}{0.1,0.1,1}
\definecolor{orange}{rgb}{0.8,0.3,0}
\definecolor{magenta}{rgb}{0.9,0.1,1}

\newcommand\nPlusOne{$N$+1}

\renewcommand\tilde[1]{\mkern1mu\widetilde{\mkern-1mu#1}}

\global\long\def\ii{\mathrm{i}}
\global\long\def\ee{\mathrm{e}}
\global\long\def\dd{\mathrm{d}}
\global\long\def\ppi{\mathrm{\pi}}
\global\long\def\tr{\mathsf{{\scriptscriptstyle T}}}
\global\long\def\Tr{\operatorname{Tr}}
\global\long\def\op#1{\operatorname{#1}}
\global\long\def\dim{\operatorname{dim}}
\global\long\def\diag{\operatorname{diag}}
\global\long\def\Lie{\mathrm{\mathscr{L}}}

\global\long\def\mfrac#1#2{\frac{\raisebox{-0.45ex}{\scalebox{0.9}{#1}}}{\raisebox{0.4ex}{\scalebox{0.9}{#2}}}}
\global\long\def\mbinom#1#2{\Big(\begin{array}{c}
 #1\\[-0.75ex]
 #2 
\end{array}\Big)}

\global\long\def\tud#1#2#3{#1{}^{#2}{}_{#3}}
\global\long\def\tdu#1#2#3{#1{}_{#2}{}^{#3}}

\global\long\def\qvf{\xi}
\global\long\def\ixA{a}
\global\long\def\ixB{b}
\global\long\def\ccVar{\mathcal{C}}

\global\long\def\lidx#1{\ ^{(#1)}\!}

\global\long\def\gSector#1{{\color{black}#1}}
\global\long\def\fSector#1{{\color{black}#1}}
\global\long\def\hSector#1{{\color{black}#1}}
\global\long\def\sSector#1{{\color{black}#1}}
\global\long\def\lSector#1{{\color{black}#1}}
\global\long\def\mSector#1{{\color{black}#1}}
\global\long\def\hrColor#1{{\color{black}#1}}
\global\long\def\VColor#1{{\color{black}#1}}
\global\long\def\KColor#1{{\color{black}#1}}
\global\long\def\KVColor#1{{\color{black}#1}}

\global\long\def\gMet{\gSector g}
\global\long\def\gSp{\gSector{\gamma}}
\global\long\def\gK{\gSector K}
\global\long\def\gE{\gSector e}
\global\long\def\gD{\gSector D}
\global\long\def\gR{\gSector R}
\global\long\def\gCS{\gSector{\Gamma}}
\global\long\def\gVse{\gSector{V_{g}}}
\global\long\def\gTse{\gSector{T_{g}}}
\global\long\def\gEinst{\gSector{G_{g}}}
\global\long\def\gRicci{\gSector{R_{g}}}
\global\long\def\gCC{\gSector{\mathcal{C}}}
\global\long\def\gCE{\gSector{\mathcal{E}}}
\global\long\def\gCD{\gSector{\nabla}}
\global\long\def\gPi{\gSector{\pi}}

\global\long\def\fMet{\fSector f}
\global\long\def\fSp{\fSector{\varphi}}
\global\long\def\fK{\fSector{\tilde{K}}}
\global\long\def\fE{\fSector m}
\global\long\def\fD{\fSector{\tilde{D}}}
\global\long\def\fR{\fSector{\tilde{R}}}
\global\long\def\fCS{\fSector{\tilde{\Gamma}}}
\global\long\def\fVse{\fSector{V_{f}}}
\global\long\def\fTse{\fSector{T_{f}}}
\global\long\def\fEinst{\fSector{G_{f}}}
\global\long\def\fRicci{\fSector{R_{f}}}
\global\long\def\fCC{\fSector{\widetilde{\mathcal{C}}}}
\global\long\def\fCE{\fSector{\widetilde{\mathcal{E}}}}
\global\long\def\fCD{\fSector{\widetilde{\nabla}}}
\global\long\def\fPi{\fSector p}
\global\long\def\fPi{\fSector{\tilde{\pi}}}

\global\long\def\gLapse{\gSector{\alpha}}
\global\long\def\gShift{\gSector{\beta}}
\global\long\def\gShiftVec{\gSector{\beta}}

\global\long\def\fLapse{\fSector{\tilde{\alpha}}}
\global\long\def\fShift{\fSector{\tilde{\beta}}}
\global\long\def\fShiftVec{\fSector{\tilde{\beta}}}

\global\long\def\gKappa{\gSector{\kappa_{g}}}
\global\long\def\gKappainv{\gSector{\kappa_{g}^{-1}}}
\global\long\def\Mg{\gSector{M_{g}^{d-2}}}

\global\long\def\fKappa{\fSector{\kappa_{f}}}
\global\long\def\fKappainv{\fSector{\kappa_{f}^{-1}}}
\global\long\def\Mf{\fSector{M_{f}^{d-2}}}

\global\long\def\grho{\gSector{\rho}}
\global\long\def\gjota{\gSector j}
\global\long\def\gJota{\gSector J}

\global\long\def\frho{\fSector{\tilde{\rho}}}
\global\long\def\fjota{\fSector{\tilde{j}}}
\global\long\def\fJota{\fSector{\tilde{J}}}

\global\long\def\grhom{\gSector{\rho^{\mathrm{m}}}}
\global\long\def\gjotam{\gSector{j^{\mathrm{m}}}}
\global\long\def\gJotam#1{\gSector{J_{#1}^{\mathrm{m}}}}

\global\long\def\frhom{\fSector{\tilde{\rho}^{\mathrm{m}}}}
\global\long\def\fjotam{\fSector{\tilde{j}^{\mathrm{m}}}}
\global\long\def\fJotam#1{\fSector{\tilde{J}_{#1}^{\mathrm{m}}}}

\global\long\def\grhob{\gSector{\rho^{\mathrm{b}}}}
 \global\long\def\gjotab{\gSector{j^{\mathrm{b}}}}
 \global\long\def\gJotab{\gSector{J^{\mathrm{b}}}}

\global\long\def\frhob{\fSector{\tilde{\rho}^{\mathrm{b}}}}
 \global\long\def\fjotab{\fSector{\tilde{j}^{\mathrm{b}}}}
 \global\long\def\fJotab{\fSector{\tilde{J}^{\mathrm{b}}}}

\global\long\def\grhoeff{\gSector{\rho_{\mathrm{eff}}}}
 \global\long\def\gjotaeff{\gSector{j_{\mathrm{eff}}}}
 \global\long\def\gJotaeff{\gSector{J_{\mathrm{eff}}}}

\global\long\def\frhoeff{\fSector{\tilde{\rho}_{\mathrm{eff}}}}
 \global\long\def\fjotaeff{\fSector{\tilde{j}_{\mathrm{eff}}}}
 \global\long\def\fJotaeff{\fSector{\tilde{J}_{\mathrm{eff}}}}

\global\long\def\gAlpha{\gSector{\alpha}}
\global\long\def\gBeta{\gSector{\beta}}
\global\long\def\gEA{\gSector A}
\global\long\def\gEB{\gSector B}
\global\long\def\fAlpha{\fSector{\tilde{\alpha}}}
\global\long\def\fBeta{\fSector{\tilde{\beta}}}
\global\long\def\fEA{\fSector{\tilde{A}}}
\global\long\def\fEB{\fSector{\tilde{B}}}

\global\long\def\sEtau{\mSector{\tau}}
\global\long\def\sESigma{\mSector{\Sigma}}
\global\long\def\sER{\mSector R}

\global\long\def\Proj{\operatorname{\perp}}
\global\long\def\gProj{\gSector{\operatorname{\perp}_{g}}}
\global\long\def\fProj{\fSector{\operatorname{\perp}_{f}}}
\global\long\def\hProj{\hSector{\operatorname{\perp}}}
\global\long\def\prho{\boldsymbol{\rho}}
\global\long\def\pjota{\boldsymbol{j}}
\global\long\def\pJota{\boldsymbol{J}}

\global\long\def\sgn{\gSector{\mathsfit{n}{\mkern1mu}}}
\global\long\def\sgD{\gSector{\mathcal{D}}}
\global\long\def\sgQ{\gSector{\mathcal{Q}}}
\global\long\def\sgV{\gSector{\mathcal{V}}}
\global\long\def\sgU{\gSector{\mathcal{U}}}
\global\long\def\sgB{\gSector{\mathcal{B}}}

\global\long\def\sfn{\fSector{\tilde{\mathsfit{n}}{\mkern1mu}}}
\global\long\def\sfD{\fSector{\widetilde{\mathcal{D}}}}
\global\long\def\sfQ{\fSector{\widetilde{\mathcal{Q}}}}
\global\long\def\sfV{\fSector{\widetilde{\mathcal{V}}}}
\global\long\def\sfU{\fSector{\widetilde{\mathcal{U}}}}
\global\long\def\sfB{\fSector{\widetilde{\mathcal{B}}}}

\global\long\def\sgW{\gSector{\mathcal{W}}}
\global\long\def\sgQU{\gSector{(\mathcal{Q\fSector{{\scriptstyle \widetilde{U}}}})}}

\global\long\def\sfW{\fSector{\tilde{\mathcal{W}}}}
\global\long\def\sfQU{\fSector{(\mathcal{\widetilde{Q}\gSector{{\scriptstyle U}}})}}

\global\long\def\hMet{\hSector h}
\global\long\def\hSp{\hSector{\chi}}
\global\long\def\hLapse{\hSector H}
\global\long\def\hShift{\hSector q}
\global\long\def\hShiftVec{\hSector q}
\global\long\def\hCC{\hSector{\bar{\mathcal{C}}}}

\global\long\def\sLs{\lSector{\hat{\Lambda}}}
\global\long\def\sLt{\lSector{\lambda}}
\global\long\def\sLtinv{\lSector{\lambda^{-1}}}
\global\long\def\sLv{\lSector v}
\global\long\def\sLp{\lSector p}
\global\long\def\sRs{\lSector{\hat{R}}}
\global\long\def\sRbar{\lSector{\bar{R}}}

\global\long\def\sI{\lSector{\hat{I}}}
\global\long\def\sEta{\lSector{\hat{\delta}}}

\global\long\def\betaSum{m^{4}{\textstyle \sum_{n}}\beta_{(n)}}
\global\long\def\betaSumL{m^{4}{\displaystyle \sum_{n}}\beta_{(n)}}

\global\long\def\signV{\,+\,}
\global\long\def\isignV{\,-\,}
\global\long\def\usignV{}
\global\long\def\uisignV{-\,}

\global\long\def\isignV{\,+\,}
\global\long\def\signV{\,-\,}
\global\long\def\uisignV{}
\global\long\def\usignV{-\,}

\global\long\def\signK{\,+\,}
\global\long\def\isignK{\,-\,}
\global\long\def\usignK{}
\global\long\def\uisignK{-\,}

\global\long\def\isignK{\,+\,}
\global\long\def\signK{\,-\,}
\global\long\def\uisignK{}
\global\long\def\usignK{-\,}

\global\long\def\signKV{\,+\,}
\global\long\def\isignKV{\,-\,}
\global\long\def\usignKV{}
\global\long\def\uisignKV{-\,}

\global\long\def\isignKV{\,+\,}
\global\long\def\signKV{\,-\,}
\global\long\def\uisignKV{}
\global\long\def\usignKV{-\,}

\global\long\def\signKV{\,+\,}
\global\long\def\isignKV{\,-\,}
\global\long\def\usignKV{+}
\global\long\def\uisignKV{-\,}

\global\long\def\isignKV{\,+\,}
\global\long\def\signKV{\,-\,}
\global\long\def\uisignKV{}
\global\long\def\usignKV{-\,}

\global\long\def\signKV{\,+\,}
\global\long\def\isignKV{\,-\,}
\global\long\def\usignKV{}
\global\long\def\uisignKV{-\,}

\global\long\def\hrD{\hrColor D}
\global\long\def\hrQ{\hrColor Q}
\global\long\def\hrn{\hrColor n}
\global\long\def\hrDn{\hrColor{Dn}}
\global\long\def\hrx{\hrColor x}

\global\long\def\hrV{\hrColor V}
\global\long\def\hrU{\hrColor U}
\global\long\def\hrVbar{\hrColor{\bar{V}}}
\global\long\def\hrWbar{\hrColor{\bar{W}}}
\global\long\def\hrSV{\hrColor S}
\global\long\def\hrUtilde{\hrColor{\tilde{U}}}
\global\long\def\hrVubar{\hrColor{\underbar{V}}}

\global\long\def\hrD{\mathsfit{D}{\mkern1mu}}

\global\long\def\CN{\gSector{\mathcal{C}}}
 \global\long\def\CNdot{\dot{\gSector{\mathcal{C}}}}
 \global\long\def\gCvec{\gSector{\mathcal{C}}}
\global\long\def\CNsm#1{\CN[#1]}

\global\long\def\CL{\fSector{\widetilde{\fSector{\mathcal{C}}}}}
 \global\long\def\CLdot{\dot{\CL}}
 \global\long\def\fCvec{\fSector{\widetilde{\fSector{\mathcal{C}}}}}
\global\long\def\CLsm#1{\CL[#1]}

\global\long\def\Ctwo{\mSector{\mathcal{C}_{2}}}
 \global\long\def\Ctwodot{\dot{\mSector{\mathcal{C}}}_{\mSector 2}}
\global\long\def\Ctwosm#1{\Ctwo[#1]}

\global\long\def\gfCvec{\mSector{\mathcal{R}}}

\global\long\def\fsm{\xi}
\global\long\def\ksm{\eta}

\global\long\def\Cbim{\mSector{\mathcal{C}_{\mathrm{b}}}}

\global\long\def\gW{\gSector{W_{g}}}
 \global\long\def\fW{\fSector{W_{f}}}

\global\long\def\gCW{\gSector{\Omega_{g}}}
 \global\long\def\fCW{\fSector{\Omega_{f}}}

\global\long\def\fWA{A}
\global\long\def\fWB{B}
\global\long\def\fWC{C}
\global\long\def\fWD{D}

\global\long\def\dt{\partial_{t}}
\global\long\def\dtp{\partial_{t} \sLp}
\global\long\def\fEtr{\fSector{m_{\mathtt{o}}}}

\global\long\def\rb#1{\left(#1\right)}
 \global\long\def\qb#1{\left[#1\right]}
 \global\long\def\cb#1{\left\lbrace #1 \right\rbrace }
 \global\long\def\qm#1{``#1''}

\global\long\def\vect#1{\operatorname{vec}\qb{#1}}

\global\long\def\gblab{\gSector{\mathrm{b}}}
\global\long\def\gmlab{\gSector{\mathrm{m}}}
\global\long\def\fblab{\fSector{\mathrm{b}}}
\global\long\def\fmlab{\fSector{\mathrm{m}}}

\fi
%%%%%%%%%%%%%%%%%%%%%%%%%%%%%%%%%%%%%%%%%%%%%%%%%%%%%%%%%%%%%%%%%%%%%%%%%%%%%

%%%%%%%%%%%%%%%%%%%%%%%%%%%%%%%%%%%%%%%%%%%%%%%%%%%%%%%%%%%%%%%%%%%%%%%%%%%%%
% Front matter
%%%%%%%%%%%%%%%%%%%%%%%%%%%%%%%%%%%%%%%%%%%%%%%%%%%%%%%%%%%%%%%%%%%%%%%%%%%%%

\newcommand{\myReleaseInfo}{~\\~} % v0.11 2019-03-22

\newcommand{\myTitle}{On the ratio of lapses in bimetric relativity}

\newcommand{\myAbstract}{The two lapse functions in the Hassan\textendash Rosen
bimetric theory are not independent. Without knowing the relation
between them, one cannot evolve the equations in the 3+1 formalism.
This work computes the ratio of lapses for the spherically symmetric
case, which is a prerequisite for numerical bimetric relativity.}

\newcommand{\myKeywords}{\bgroup\small Modified gravity, Bimetric
relativity, Ghost-free bimetric theory\egroup}

\title{\myTitle}

\author{Mikica Kocic,}

\author{Anders Lundkvist, and}

\author{Francesco Torsello}

\affiliation{
  Department of Physics \& The Oskar Klein Centre,\\
  Stockholm University, AlbaNova University Centre,
  SE-106 91 Stockholm
}

\email{mikica.kocic@fysik.su.se}

\email{anders.lundkvist@fysik.su.se}

\email{francesco.torsello@fysik.su.se}

\hypersetup{
  pdftitle=\myTitle,
  pdfauthor=Mikica Kocic and Anders Lundkvist and Francesco Torsello,
  pdfsubject=Hassan-Rosen ghost-free bimetric theory,
  pdfkeywords={Modified gravity, Ghost-free bimetric theory}
}

%\notoctrue
\emitFrontMatter%\vspace{-3ex}

%%%%%%%%%%%%%%%%%%%%%%%%%%%%%%%%%%%%%%%%%%%%%%%%%%%%%%%%%%%%%%%%%%%%%%%%%%%%%
% Main matter
%%%%%%%%%%%%%%%%%%%%%%%%%%%%%%%%%%%%%%%%%%%%%%%%%%%%%%%%%%%%%%%%%%%%%%%%%%%%%

%\graphicspath{{../}}

\section{Introduction}

This work establishes the relationship between the two lapse functions
in the Hassan-Rosen (HR) bimetric theory when the equations are reduced
to the spherically symmetric case. The HR theory \cite{Hassan:2011zd,Hassan:2011ea,Hassan:2017ugh,Hassan:2018mbl}
is a nonlinear theory of two interacting classical spin-2 fields.
It is closely related to de Rham\textendash Gabadadze\textendash Tolley
(dRGT) massive gravity, where one of the metrics is frozen and taken
to be a nondynamical fiducial metric \cite{deRham:2010ik,deRham:2010kj,Hassan:2011hr}.
Comprehensive reviews of these theories can be found in \cite{Schmidt-May:2015vnx,deRham:2014zqa}. 

As shown in \cite{Hassan:2018mbl,Alexandrov:2012yv}, the ratio of
the two lapses in the HR theory is a function of the dynamical variables
only. Without exactly knowing their relation in 3+1 formalism, one
cannot set up the initial value problem and evolve the equations of
motion in numerical bimetric relativity. Here we evaluate the ratio
of the lapses for the case of spherical symmetry. The calculation
is based on the evolution equations in standard 3+1 form \cite{Kocic:2018ddp}.

\subsubsection*{Background}

\label{ssec:background}

\begin{figure}[h]
\noindent \begin{centering}
\includegraphics[scale=0.8]{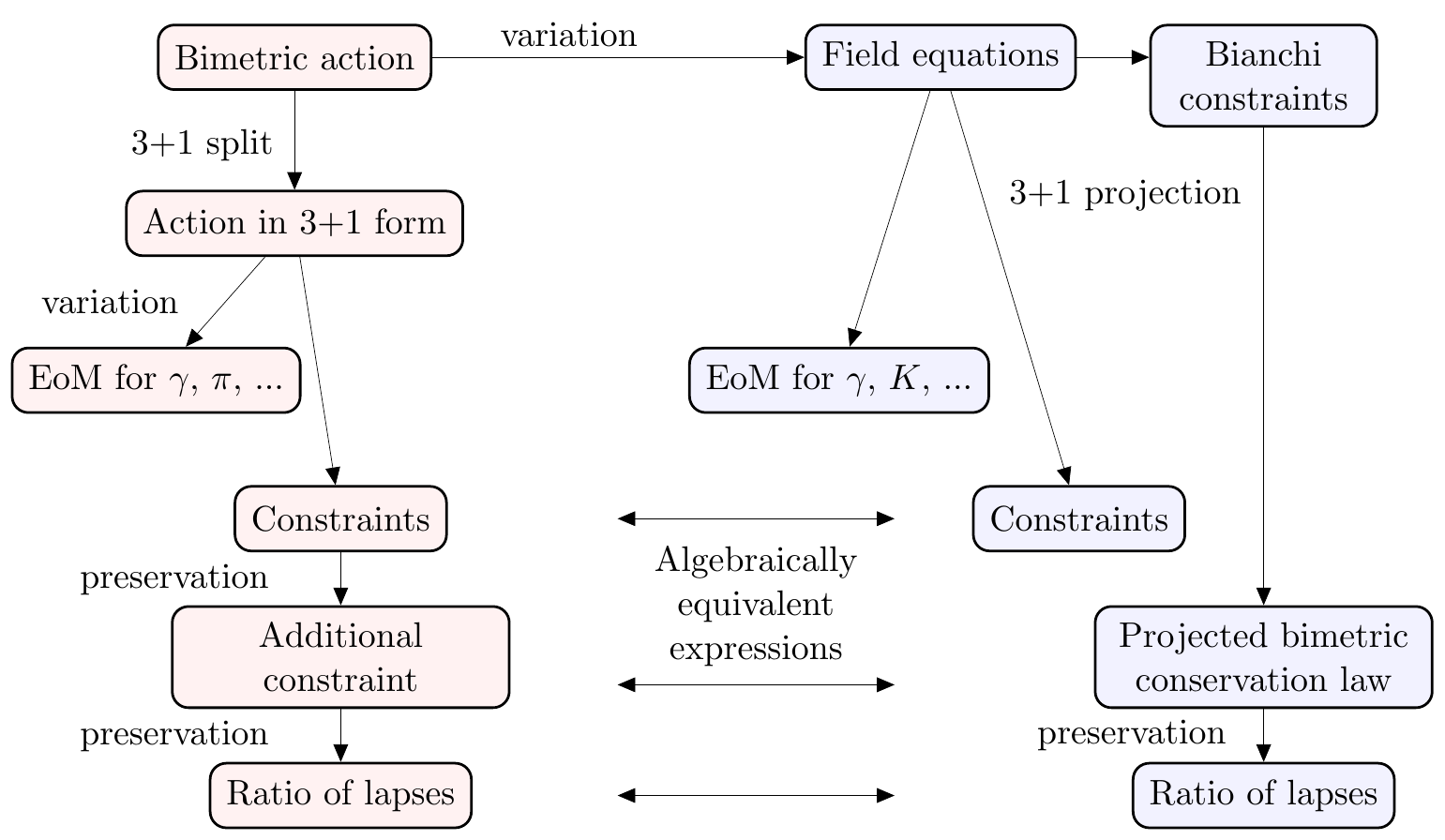}
\par\end{centering}
\caption{\label{fig:comp}Comparison between two approaches: (i) the canonical
analysis of the constraints, and (ii) the 3+1 projection of the bimetric
field equations (together with the Bianchi constraints).}
\end{figure}

Similarly to general relativity \cite{Arnowitt:1962hi,York:1979aa},
the kinematical and dynamical parts of the metric fields in bimetric
relativity can be isolated using the 3+1 formalism. However, there
are two seemingly different paths in attacking the problem. One approach
begins with the 3+1 split at the level of the action. This path is
suitable for the Hamiltonian formulation and the canonical analysis
of constraints \cite{Hassan:2018mbl}. The other approach starts from
the 3+1 projection of the field equations \cite{Kocic:2018ddp}. This
is more in line with the initial value formulation of the theory.
The comparison of these two approaches is shown in Figure \ref{fig:comp}.

\noindent These procedures should and do yield equivalent results.
The canonical analysis provides a fundamental view on the structure
and the relation between the constraints. On the other hand, a direct
3+1 projection of the field equations is less involved (i.e., faster),
and more suitable as the starting point towards numerical bimetric
relativity. In the rest of the paper we traverse both paths in Figure
\ref{fig:comp}. More specifically, the canonical analysis is used
to prove that the ratio of lapses only depends on the dynamical fields
in the most general case. The actual calculations in spherical symmetry
will be based on the HR field equations in standard 3+1 form.

The rest of this paper is structured as follows. We begin by stating
the action, the field equations, and the 3+1 decomposition of both
the metrics and the square root in the potential. We then investigate
the relationship between the lapses using the constraint analysis,
and quote the basic equations that are used as a starting point. Subsection
\ref{sec:cons-law} states the bimetric conservation law (the so-called
secondary constraint) for the case of spherical symmetry, and establishes
the evolution equation for the relative shift between the two metrics,
called the `separation parameter'. Subsection \ref{sec:ratio} presents
the actual derivation of the ratio of lapses. The paper ends with
a short discussion of the result.

\paragraph*{Action.}

The Hassan-Rosen action reads \cite{Hassan:2011zd},
\begin{align}
\mathcal{S} & =\int\dd^{4}x\sqrt{-\gMet}\,\bigg[\frac{1}{2\gKappa}\,\gRicci+\mathcal{\gSector L}_{\gMet}^{\gmlab}\bigg]+\int\dd^{4}x\sqrt{-\fMet}\,\bigg[\frac{1}{2\fKappa}\,\fRicci+\mathcal{\fSector L}_{\fMet}^{\fmlab}\bigg]\nonumber \\
 & \qquad\qquad\signV m^{4}\int\dd^{4}x\sqrt{-\gMet}\,\sum_{n=0}^{4}\beta_{(n)}e_{n}(\sqrt{\gMet^{-1}\fMet}),\label{eq:bim-action}
\end{align}
where $\gRicci$ and $\fRicci$ are the Ricci scalars of $\gMet$
and $\fMet$, respectively, $\gKappa$ and $\fKappa$ are the Einstein
gravitational constants for the two sectors, and $\beta_{(n)}$ are
free parameters which are dimensionless and scaled by $m$. The respective
Lagrangian densities of the two matter sectors are denoted as $\mathcal{\gSector L}_{\gMet}^{\gmlab}$
and $\mathcal{\fSector L}_{\fMet}^{\fmlab}$. The ghost-free bimetric
interactions are specifically constructed using the elementary symmetric
polynomials $e_{n}$ in terms of the principal square root matrix
$(\gMet^{-1}\fMet)^{1/2}$ that represents a (1,1) tensor field \cite{Hassan:2017ugh}.

\paragraph*{Field equations.}

The bimetric field equations can be written,
\begin{equation}
\gEinst=\gKappa(\gVse+\gTse),\qquad\fEinst=\fKappa(\fVse+\fTse),\label{eq:bim-fe}
\end{equation}
where $\gEinst$ and $\fEinst$ are the Einstein tensors of $\gMet$
and $\fMet$, $\gTse$ and $\fTse$ are the stress\textendash energy
tensors of the matter fields each minimally coupled to a different
sector, and $\gVse$ and $\fVse$ are the stress\textendash energy
contributions of the ghost-free bimetric potential, also known as
the bimetric stress\textendash energy tensors.

\paragraph*{The 3+1 decomposition.}

We assume that the metrics are in the usual 3+1 form,\bSe\label{eq:gf-Met}
\begin{align}
\gMet & =-\gLapse^{2}\dd t^{2}+\gSp_{ij}\big(\dd x^{i}+\gShift{}^{i}\dd t\big)\big(\dd x^{j}+\gShift{}^{j}\dd t\big),\\
\fMet & =-\fLapse^{2}\dd t^{2}+\fSp_{ij}\big(\dd x^{i}+\fShift{}^{i}\dd t\big)\big(\dd x^{j}+\fShift{}^{j}\dd t\big),
\end{align}
\eSe where $\gSp_{ij}$ and $\fSp_{ij}$ are the spatial metrics,
$\gShift{}^{i}$ and $\fShift{}^{i}$ are the shift vectors, and $\gLapse$
and $\fLapse$ are the lapse functions. 

\paragraph*{The square root.}

The ghost-free bimetric interactions are specifically constructed
in terms of the square root matrix $(\gMet^{-1}\fMet)^{1/2}$. The
chief condition is the existence and uniqueness of the real square
root \cite{Hassan:2017ugh}. Upon 3+1 decomposition, the condition
can be expressed in terms of the affine related variables $\sgn^{i}$
and $\sfn^{i}$ via the \emph{mean shift vector} $\hShift^{i}$,
\begin{equation}
\hShift^{i}\coloneqq\gShift^{i}-\gLapse\sgn^{j}=\fShift^{i}+\fLapse\sfn^{i}.\label{ndef}
\end{equation}
These variables are connected through $\sgn^{i}=\tud{\hrD}ij\sfn^{j}$,
or in matrix notation,
\begin{equation}
\sgn=\mathsf{\hrD}\sfn,\qquad\sgn=\gE^{-1}\sLv,\qquad\sfn=\fE^{-1}\sLv,
\end{equation}
where $\hrD$ is defined in \cite{Hassan:2011tf}, $\gE$ and $\fE$
are the symmetrized spatial vielbeins of $\gSp$ and $\fSp$, while
$\sLv$ is the boost vector of the Lorentz transformation that is
used to symmetrize the vielbeins ensuring the reality of the square
root \cite{Hassan:2014gta,Kocic:2018ddp}. Geometrically, these shift-like
vectors encode the separation (a relative shift) between the two metrics,
which can equally be parametrized by $\sLv$. In other words, the
variables $\sgn$, $\sfn$, and $\sLv$ are ``three'' sides of the
same coin. Note that $\hrD$ does not depend on the two lapse functions,
and that $\mathsf{\hrD}\ne\gE^{-1}\fE$ because of the overall local
Lorentz invariance that was used to symmetrize the vielbeins.

\subsection{Canonical analysis of the constraints}

In the following two subsections we briefly summarize the constraint
analysis of \cite{Hassan:2018mbl}. In order to isolate the constraints,
we need to eliminate one of the shift vectors, for instance $\gShift^{i}$,
in terms of one of the new variables, for example $\sfn^{i}$, using
\eqref{ndef},
\begin{equation}
\gShift^{i}=\fShift^{i}+\fLapse\sfn^{i}+\gLapse\sgn^{j}=\fShift^{i}+\fLapse\sfn^{i}+\gLapse\tud{\hrD}ij\sfn^{j}.
\end{equation}
Note that this choice is arbitrary. For this particular choice, the
Lagrangian is linear in $\gLapse$, $\fLapse$ and $\fShift^{i}$,
and can be written, 
\begin{equation}
\mathcal{L}=\gPi^{ij}\dot{\gSp}_{ij}+\fPi^{ij}\dot{\fSp}_{ij}+\gLapse\CN+\fLapse\CL+\fShift^{i}\gfCvec_{i},\label{HRLag}
\end{equation}
where $\gPi^{ij}$ and $\fPi^{ij}$ denote the canonical momenta of
$\gSp_{ij}$ and $\fSp_{ij}$, respectively. The quantities $\CN$,
$\CL$, and $\gfCvec_{i}$ are defined in \eqref{eq:g-cc} and \eqref{eq:h-cc-mixed}.
Variation with respect to the lapse functions and the shift vector
gives rise to the constraints, 
\begin{equation}
\CN=0,\qquad\CL=0,\qquad\gfCvec_{i}=0.\label{Lagconst}
\end{equation}
These equations depend on $\sfn^{i}$ and varying the Lagrangian with
respect to $\sfn^{i}$ yields its equations of motion,
\begin{equation}
\gCvec_{i}=0,\label{neom}
\end{equation}
where $\gCvec_{i}$ is defined in \eqref{eq:g-cc-vector}. Equation
\eqref{neom} can in principle be solved for $\sfn^{i}$ in terms
of the dynamical variables.%
\begin{comment}
An explicit solution is only known for some special cases, but such
a solution is not needed here. See appendix \ref{app:dtp} for more
details.
\end{comment}
{} After imposing that solution on \eqref{Lagconst}, those equations
depend only on the dynamical variables.

Since the constraints are valid at all times, their time derivatives
must vanish on the constraint surface. This can be used to find an
additional constraint. The time derivative of a quantity is determined
by its Poisson bracket with the Hamiltonian, 
\begin{equation}
H=-\int\dd^{3}x\left(\gLapse\CN+\fLapse\CL+\fShift^{i}\gfCvec_{i}\right).\label{H}
\end{equation}
From this we see that the time derivatives of the constraints involve
Poisson brackets of the constraints with each other. These brackets
are computed in \cite{Hassan:2018mbl} and read, 
\begin{align}
\{\gfCvec_{i}(x),\gfCvec_{j}(y)\} & =-\left[\gfCvec_{j}(x)\frac{\partial}{\partial x^{i}}\delta^{3}(x-y)-\gfCvec_{i}(y)\frac{\partial}{\partial y^{j}}\delta^{3}(x-y)\right]\text{,}\label{RiRj}\\
\{\CL(x),\gfCvec_{i}(y)\} & =-\CL(y)\frac{\partial}{\partial x^{i}}\delta^{3}(x-y)\text{,}\label{CLRi}\\
\{\CL(x),\CL(y)\} & =-\left[\fSp^{ij}(x)\gfCvec_{j}(x)\frac{\partial}{\partial x^{i}}\delta^{3}(x-y)-\fSp^{ij}(y)\gfCvec_{j}(y)\frac{\partial}{\partial y^{i}}\delta^{3}(x-y)\right]\text{,}\label{CLCL}\\
\{\CN(x),\gfCvec_{i}(y)\} & =-\CN(y)\frac{\partial}{\partial x^{i}}\delta^{3}(x-y)\text{,}\label{CNRi}\\
\{\CN(x),\CN(y)\} & =-\left[\CN(x)\sgn^{i}(x)\frac{\partial}{\partial x^{i}}\delta^{3}(x-y)-\CN(y)\sgn^{i}(y)\frac{\partial}{\partial y^{i}}\delta^{3}(x-y)\right]\text{,}\label{CNCN}\\
\{\CL(x),\CN(y)\} & =\Ctwo(x)\delta^{3}(x-y)\text{.}\label{CLCN}
\end{align}
Here, the expression for $\Ctwo$ is a function of the phase space
variables defined in \eqref{eq:bim-prop-constr}. Note that all of
the brackets, except for \eqref{CLCN}, vanish upon imposing the constraints
\eqref{Lagconst}. Therefore, it follows that, 
\begin{equation}
\CNdot=\{\CN,H\}\approx\fLapse\Ctwo,\qquad\CLdot=\{\CL,H\}\approx-\gLapse\Ctwo,\qquad\dot{\gfCvec}_{i}=\{\gfCvec_{i},H\}\approx0,\label{constraintsdot}
\end{equation}
where the symbol $\approx$ denotes weak equality, that is, equality
on the constraint surface. Since $\gLapse$ and $\fLapse$ are nonzero
and all expressions in \eqref{constraintsdot} vanish on the constraint
surface, this means that we must have, 
\begin{equation}
\Ctwo=0,\label{C2}
\end{equation}
which provides us with an additional constraint.

\subsection{Ratio of lapses in the canonical analysis}

\label{cratio}

Since the constraint \eqref{C2} must be valid at all times, it is
necessary that $\Ctwodot\approx0$, where $\approx$ now denotes equality
on the surface of all six constraints \eqref{Lagconst} and \eqref{C2}.
Here we show that the requirement $\Ctwodot\approx0$ gives a linear
relation between the two lapses for the most general case. The time
derivative is given by, 
\begin{align}
\Ctwodot(x) & =\{\Ctwo(x),H\}\nonumber \\
 & =-\int\dd^{3}y\,\Big(\gLapse(y)\{\Ctwo(x),\CN(y)\}+\fLapse(y)\{\Ctwo(x),\CL(y)\}+\fLapse^{i}(y)\{\Ctwo(x),\gfCvec_{i}(y)\}\Big).\label{C2dot}
\end{align}
We now consider each of the Poisson brackets that appear in this expression.
In order to compute them we use, 
\begin{equation}
\Ctwo(x)=\int\dd^{3}z\,\{\CL(x),\CN(z)\},
\end{equation}
which follows from \eqref{CLCN}. We also use the Jacobi identity,
$\{\{A,B\},C\}=\{A,\{B,C\}\}-\{B,\{A,C\}\}$. The final Poisson bracket
in \eqref{C2dot} can then be written, 
\begin{align}
\{\Ctwo(x),\gfCvec_{i}(y)\} & =\int\dd^{3}z\,\{\{\CL(x),\CN(z)\},\gfCvec_{i}(y)\}=-\Ctwo(y)\frac{\partial}{\partial x^{i}}\delta^{3}(x-y),\label{C2Ri}
\end{align}
where \eqref{CLRi}, \eqref{CNRi}, and \eqref{CLCN} have been used
(see appendix \ref{app:poiss} for the details). Note that this means
that this bracket vanishes on the constraint surface.

We turn our attention to the second bracket in \eqref{C2dot}. It
can be written, 
\begin{align}
\{\Ctwo(x),\CL(y)\} & =\int\dd^{3}z\,\{\{\CL(x),\CN(z)\},\CL(y)\}\nonumber \\
 & =\int\dd^{3}z\,\left(\{\CL(x),\{\CN(z),\CL(y)\}\}-\{\CN(z),\{\CL(x),\CL(y)\}\}\right).\label{eq:C2CL}
\end{align}
From \eqref{CLCL} and \eqref{CNRi}, it follows that the second term
vanishes; hence, 
\begin{align}
\{\Ctwo(x),\CL(y)\} & =\int\dd^{3}z\,\{\CL(x),\{\CN(z),\CL(y)\}\}=-\int\dd^{3}z\,\{\CL(x),\Ctwo(y)\}\delta^{3}(y-z)\nonumber \\
 & =-\{\CL(x),\Ctwo(y)\}=\{\Ctwo(y),\CL(x)\},\label{C2CLsimp}
\end{align}
where we have used \eqref{CLCN}. We see that this bracket is symmetric
with respect to interchange of $x$ and $y$. The first bracket in
\eqref{C2dot} can be dealt with in a similar way, 
\begin{align}
\{\Ctwo(x),\CN(y)\} & =\int\dd^{3}z\,\{\{\CL(x),\CN(z)\},\CN(y)\}=\int\dd^{3}z\,\{\{\CL(z),\CN(x)\},\CN(y)\}\nonumber \\
 & =\int\dd^{3}z\,\left(\{\CL(z),\{\CN(x),\CN(y)\}\}-\{\CN(x),\{\CL(z),\CN(y)\}\}\right).\label{eq:C2CN}
\end{align}
It follows from \eqref{CNCN} and \eqref{CLCN} that the first term
vanishes weakly, implying,
\begin{align}
\{\Ctwo(x),\CN(y)\} & \approx-\int\dd^{3}z\{\CN(x),\{\CL(z),\CN(y)\}\}=-\int\dd^{3}z\{\CN(x),\Ctwo(z)\}\delta^{3}(z-y)\label{C2CNsimp}\\
 & =-\{\CN(x),\Ctwo(y)\}=\{\Ctwo(y),\CN(x)\}.
\end{align}
To summarize, we have, 
\begin{align}
\{\Ctwo(x),\CL(y)\} & =\{\Ctwo(y),\CL(x)\},\label{C2CLsym}\\
\{\Ctwo(x),\CN(y)\} & \approx\{\Ctwo(y),\CN(x)\}.\label{C2CNsym}
\end{align}
In order to compute these brackets, define the smeared constraints,
\begin{equation}
\Ctwosm{\fsm}\coloneqq\int\dd^{3}x\,\fsm(x)\Ctwo(x),\quad\CLsm{\ksm}\coloneqq\int\dd^{3}y\,\ksm(y)\CL(y),\quad\CNsm{\ksm}\coloneqq\int\dd^{3}y\,\ksm(y)\CN(y),
\end{equation}
where $\fsm$ and $\ksm$ are arbitrary localized smoothing functions.
It follows that, 
\begin{equation}
\{\Ctwosm{\fsm},\CLsm{\ksm}\}=\iint\dd^{3}x\dd^{3}y\,\fsm(x)\ksm(y)\,\{\Ctwo(x),\CL(y)\},
\end{equation}
which means that if we compute $\{\Ctwosm{\fsm},\CLsm{\ksm}\}$, we
can extract $\{\Ctwo(x),\CL(y)\}$. The most general expression for
$\{\Ctwosm{\fsm},\CLsm{\ksm}\}$ is \cite{Alexandrov:2012yv}, 
\begin{equation}
\{\Ctwosm{\fsm},\CLsm{\ksm}\}=\int\dd^{3}z\,\left(\fsm\ksm\fWA+\fsm\,\fWB^{i}\partial_{i}\ksm+\ksm\,\fWC^{i}\partial_{i}\fsm\right),\label{general}
\end{equation}
where $\fWA$, $\fWB^{i}$ and $\fWC^{i}$ are some functions of the
phase space variables. From \eqref{C2CLsym} it follows that this
expression should be symmetric with respect to interchanges of $\fsm$
and $\ksm$. In order for this to be the case, we must have $\fWC^{i}=\fWB^{i}$.
Consequently, 
\begin{align}
\{\Ctwosm{\fsm},\CLsm{\ksm}\} & =\int\dd^{3}z\,\left(\fsm\ksm\fWA+\fsm\,\fWB^{i}\partial_{i}\ksm+\ksm\,\fWB^{i}\partial_{i}\fsm\right)=\int\dd^{3}z\,\fsm\ksm\left(\fWA-\partial_{i}\fWB^{i}\right)\nonumber \\
 & =\int\dd^{3}z\,\fsm\ksm\,(-\fCW),\label{eq:C2CLs}
\end{align}
where in the second step we use integration by parts. Here we have
defined $\fCW\coloneqq-(\fWA-\partial_{i}\fWB^{i})$. Hence, we obtain,
\begin{equation}
\{\Ctwo(x),\CL(y)\}=-\fCW(x)\delta^{3}(x-y).\label{C2CL}
\end{equation}
We can use identical reasoning when computing $\{\Ctwosm{\fsm},\CNsm{\ksm}\}$.
Using \eqref{C2CNsym} we get, 
\begin{equation}
\{\Ctwo(x),\CN(y)\}\approx-\gCW(x)\delta^{3}(x-y).\label{C2CN}
\end{equation}
The functions $\gCW$ and $\fCW$ depend on the phase space variables.
We can now plug the expressions \eqref{C2Ri}, \eqref{C2CL} and \eqref{C2CN}
into \eqref{C2dot} to arrive at, 
\begin{align}
\Ctwodot(x) & \approx\int\dd^{3}y\,\left(\gLapse(y)\gCW(x)\delta^{3}(x-y)+\fLapse(y)\fCW(x)\delta^{3}(x-y)+\fShift^{i}(y)\Ctwo(y)\frac{\partial}{\partial x^{i}}\delta^{3}(x-y)\right)\nonumber \\
 & =\gLapse(x)\gCW(x)+\fLapse(x)\fCW(x)+\frac{\partial}{\partial x^{i}}\left(\fShift^{i}(x)\Ctwo(x)\right).
\end{align}
Note that the terms involving the shift vanish weakly, being proportional
to $\Ctwo$ or its spatial derivative. As mentioned, $\Ctwodot\approx0$,
which implies, 
\begin{equation}
\gLapse\gCW+\fLapse\fCW\approx0.\label{lapseeq}
\end{equation}
This yields the ratio of lapses, 
\begin{equation}
\gLapse=\mSector W\fLapse,\qquad\mSector W\coloneqq-\frac{\fCW}{\gCW}.
\end{equation}
In conclusion, the requirement $\Ctwodot\approx0$ gives a linear
relation between the two lapses.

\subsection{Bimetric field equations in standard 3+1 form}

\label{ssec:basic-eqs}

Here we highlight the second approach in Figure \ref{fig:comp}, where
the evolution and the constraint equations are obtained by projection
of the bimetric field equations. After the projection, the standard
3+1 evolution equations are \cite{Kocic:2018ddp},\bSe\label{eq:g-evol}
\begin{alignat}{2}
\partial_{t}\gSp_{ij} & =\Lie_{\gShiftVec}\gSp_{ij} &  & \signK2\gLapse\gK_{ij},\\
\partial_{t}\gK_{ij} & =\Lie_{\gShiftVec}\gK_{ij} &  & \signK\gD_{i}\gD_{j}\gLapse\isignK\gLapse\,\big[\gR_{ij}-2\gK_{ik}\gK^{k}{}_{j}+\gK\gK_{ij}\big]\nonumber \\
 &  &  & \signK\gLapse\gKappa\Big\{\,\gSp_{ik}\tud{\gJota}kj-\mfrac 12\gSp_{ij}(\tud{\gJota}kk-\grho)\,\Big\},\\
\partial_{t}\fSp_{ij} & =\Lie_{\fShiftVec}\fSp_{ij} &  & \signK2\fLapse\fK_{ij},\\
\partial_{t}\fK_{ij} & =\Lie_{\fShiftVec}\fK_{ij} &  & \signK\fD_{i}\fD_{j}\fLapse\isignK\fLapse\,\big[\fR_{ij}-2\fK_{ik}\fK^{k}{}_{j}+\fK\fK_{ij}\big]\nonumber \\
 &  &  & \signK\fLapse\fKappa\Big\{\,\fSp_{ik}\tud{\fJota}kj-\mfrac 12\fSp_{ij}(\tud{\fJota}kk-\frho)\,\Big\},
\end{alignat}
\eSe where $\Lie$ denotes a Lie derivative, $\gK_{ij}$ and $\fK_{ij}$
are the extrinsic curvatures in the two sectors, $\gR_{ij}$ and $\fR_{ij}$
are the spatial Ricci tensors, and $\gD_{i}$ and $\fD_{i}$ are the
spatial covariant derivatives compatible with $\gSp_{ij}$ and $\fSp_{ij}$,
respectively.\footnote{Note that $\gPi^{ij}=\gKappa^{-1}\sqrt{\gSp}(\gK^{ij}-\gK\,\gSp^{ij})$
and $\fPi^{ij}=\fKappa^{-1}\sqrt{\fSp}(\fK^{ij}-\fK\,\fSp^{ij})$.
The indices are raised/lowered by the metric in the respective sector.} The variables $\grho,\gjota_{i},\tud{\gJota}ij,\frho,\fjota_{i},$
and $\tud{\fJota}ij$ denote the normal, tangential and spatial projections
of the effective stress\textendash energy tensors in the two sectors,
which include both the contributions from matter and the bimetric
potential. These can be split into the bimetric contribution denoted
by upper label ``b'' and the matter contribution denoted by upper
label ``m'', for instance, $\grho=\grho^{\gblab}+\grho^{\gmlab}$.

The constraint equations are,\bSe\label{eq:g-cc}\bgroup
\begin{alignat}{3}
\gKappa\CN/\sqrt{\gSp} & =\gR+\gK^{2}-\gK_{ij}\gK^{ij} & \, & \,- & \,2\gKappa\,\grho & =0,\label{eq:g-cc-scalar}\\
\fKappa\CL/\sqrt{\fSp} & =\fR+\fK^{2}-\fK_{ij}\fK^{ij} & \, & \,- & \,2\fKappa\,\frho & \,=0,\label{eq:f-cc-scalar}\\
\gKappa\gCvec_{i}/\sqrt{\gSp} & =\gD_{k}\gK^{k}{}_{i}-\gD_{i}\gK & \, & \signK & \gKappa\,\gjota_{i} & =0,\label{eq:g-cc-vector}\\
\fKappa\fCvec_{i}/\sqrt{\fSp} & =\fD_{k}\fK^{k}{}_{i}-\fD_{i}\fK & \, & \signK & \fKappa\,\fjota_{i} & =0,\label{eq:f-cc-vector}
\end{alignat}
\egroup\eSe Note that the constraints for the mixed projections (\ref{eq:g-cc-vector})\textendash (\ref{eq:f-cc-vector})
are algebraically coupled by $\sqrt{\gSp}\,\gjota_{i}^{\gblab}+\sqrt{\fSp}\,\fjota_{i}^{\fblab}=0$,
which yields,
\begin{align}
\gfCvec_{i} & \coloneqq\sqrt{\gSp}\Big\{\gKappainv\big(\gD_{k}\gK^{k}{}_{i}-\gD_{i}\gK\big)\signK\gjota_{i}^{\gmlab}\Big\}+\,\sqrt{\fSp}\Big\{\fKappainv\big(\fD_{k}\fK^{k}{}_{i}-\fD_{i}\fK\big)\signK\fjota_{i}^{\fmlab}\Big\}=0,\label{eq:h-cc-mixed}
\end{align}
where $\gjota_{i}^{\gmlab}$ and $\fjota_{i}^{\fmlab}$ are the tangential
projections of the matter stress\textendash energy tensors only. The
last equation can be used as a replacement for either (\ref{eq:g-cc-vector})
or (\ref{eq:f-cc-vector}), where the other becomes the equation of
motion for $\sLv$ (or $\sgn$ or $\sfn$). The contributions of the
bimetric stress\textendash energy tensors can be expressed in terms
of the following set of 3+1 variables,
\begin{equation}
\{\sgn,\sgD,\sgB,\sgV,\sgU,\sgQ,\sfn,\sfD,\sfB,\sfV,\sfU,\sfQ,\sLt\}.\label{eq:Np1-vars-1}
\end{equation}
These variables, quoted in appendix \ref{app:bim-vars}, do not depend
on the mean shift $\hShift^{i}$ in (\ref{ndef}) or the lapses $\gLapse$
and $\fLapse$. 

\paragraph*{The additional constraint.}

In the 3+1 decomposition, we also have to assume specifically that
\cite{Gourgoulhon:2012trip}:
\begin{enumerate}[label={(\roman*)},itemsep=-0.7ex]
\item the matter conservation laws $\gCD_{\mu}\tud{\gTse}{\mu}{\nu}=0$
and $\fCD_{\mu}\tud{\fTse}{\mu}{\nu}=0$ hold, and that
\item the conservation law for the bimetric potential $\gCD_{\mu}\tud{\gVse}{\mu}{\nu}=0$
holds.\footnote{Note that $\sqrt{-\gMet}\,\gCD_{\mu}\tud{\gVse}{\mu}{\nu}+\sqrt{-\fMet}\,\fCD_{\mu}\tud{\fVse}{\mu}{\nu}=0$;
see \cite{Damour:2002ws}.}
\end{enumerate}
The projection of $\gCD_{\mu}\tud{\gVse}{\mu}{\nu}=0$ gives the 3+1
form of the conservation law for the ghost-free bimetric potential
\cite{Kocic:2018ddp},
\begin{align}
\Cbim & =\Ctwo/\sqrt{\gSp}=\tud{\sgU}ij\Big(\gD_{i}\sgn^{j}\signK\tud{\gK}ji\Big)+\tud{\sfU}ij\Big(\fD_{i}\sfn^{j}\isignK\tud{\fK}ji\Big)-\gD_{i}\big[\tud{\sgU}ij\sgn^{j}\big]=0,\label{eq:bim-prop-constr}
\end{align}
which is equivalent to the additional constraint (\ref{C2}) obtained
using the Hamiltonian formalism \cite{Hassan:2018mbl}. Note again
that the existence of this constraint is necessary for removing the
unphysical (ghost) modes.

\section{Ratio of lapses in spherical symmetry}

\label{sec:main}

In this section we specialize to the bimetric sectors that share the
same spherical symmetry \cite{Torsello:2017ouh}. After stating the
basic equations in standard 3+1 form, we go through the derivation
of the projected bimetric conservation law and establish the evolution
equation for the separation parameter $\sLp$. In the end, starting
from $\partial_{t}\Cbim=0$ we find the ratio between two lapse function
by using the equations of motion and the constraints.

\subsection{Basic equations }

\label{sec:ssym}

The general form of the metrics in spherical polar coordinates reads,\bSe\label{eq:ssym-gf}
\begin{align}
\gMet & =-\gAlpha^{2}\dd t^{2}+\gEA^{2}(\dd r+\gBeta\,\dd t)^{2}+\gEB^{2}(\dd\theta^{2}+\sin^{2}\theta\,\dd\phi^{2}),\\
\fMet & =-\fAlpha^{2}\dd t^{2}+\fEA^{2}(\dd r+\fBeta\,\dd t)^{2}+\fEB^{2}(\dd\theta^{2}+\sin^{2}\theta\,\dd\phi^{2}),
\end{align}
\eSe where, from now on, $\gAlpha$ and $\fAlpha$ denote the lapse
functions, and $(\gEA,\gEB,\fEA,\fEB)$ denote the nontrivial components
of the spatial vielbeins. Note that the metrics are not gauge fixed;
this is necessary to determine the general form of the ratio of lapses.
The radial components of the shifts are parametrized by the mean shift
$\hShift$ and the radial separation $\sLp$,
\begin{equation}
\gBeta\coloneqq\hShift+\gAlpha\gEA^{-1}\sLv,\quad\fBeta\coloneqq\hShift-\fAlpha\fEA^{-1}\sLv,\quad\sLv\coloneqq\sLp\sLtinv,\quad\sLt\coloneqq(1+\sLp^{2})^{1/2}.
\end{equation}
In addition, we have the components of the extrinsic curvatures,
\begin{equation}
\gK_{1}\eqqcolon\tud{\gK}rr,\quad\gK_{2}\eqqcolon\tud{\gK}{\theta}{\theta}=\tud{\gK}{\phi}{\phi},\quad\fK_{1}\eqqcolon\tud{\fK}rr,\quad\fK_{2}\eqqcolon\tud{\fK}{\theta}{\theta}=\tud{\fK}{\phi}{\phi}.
\end{equation}
All these variables are functions of $(t,r)$ to be solved for. The
3+1 variables from (\ref{eq:Np1-vars-1}) for the case of spherical
symmetry are given in appendix \ref{app:ssym-vars}. 

To simplify expressions, we introduce $\sER\coloneqq\fEB/\gEB$ and
define, 
\begin{equation}
\left\langle \sER\right\rangle _{k}^{n}\coloneqq\usignV m^{4}\sum_{l=0}^{n}\mbinom nl\beta_{(l+k)}\sER^{l},\quad\ \left\langle \sER\right\rangle _{k}^{n}=\left\langle \sER\right\rangle _{k}^{n-1}+\sER\left\langle \sER\right\rangle _{k+1}^{n-1},\ \left\langle \sER\right\rangle _{k}^{0}=\usignV m^{4}\beta_{(k)}.
\end{equation}
This way, all the $\beta_{(k)}$-parameters are kept inside $\left\langle \sER\right\rangle _{k}^{n}$. 

The projections of the effective (total) stress\textendash energy
tensor in the $\gMet$-sector are denoted by $\grho$, $\gjota_{r}$,
$\gJota_{1}\coloneqq\tud{\gJota}rr$, $\gJota_{2}\coloneqq\tud{\gJota}{\theta}{\theta}=\tud{\gJota}{\phi}{\phi}$,
and $\gJota=\gJota_{1}+2\gJota_{2}$. Similar expressions are defined
in the $\fMet$-sector. The nonzero components of the projections
of the bimetric stress\textendash energy tensors $V_{\gMet}$ and
$V_{\fMet}$ are given (\ref{eq:ssym-g-se}) and (\ref{eq:ssym-f-se})
in appendix \ref{app:ssym-vars}.

The scalar constraints (\ref{eq:g-cc-scalar})\textendash (\ref{eq:f-cc-scalar})
are,\bSe\label{eq:ssym-cc}\vspace{-0.5ex}
\begin{align}
(2\gK_{1}+\gK_{2})\gK_{2}+\frac{1}{\gEA^{2}}\bigg(\frac{\gEA^{2}}{\gEB^{2}}+2\frac{\partial_{r}\gEA}{\gEA}\frac{\partial_{r}\gEB}{\gEB}-\frac{(\partial_{r}\gEB)^{2}}{\gEB^{2}}-2\frac{\partial_{r}^{2}\gEB}{\gEB}\bigg) & =\gKappa\grho,\\
(2\fK_{1}+\fK_{2})\fK_{2}+\frac{1}{\fEA^{2}}\bigg(\frac{\fEA^{2}}{\fEB^{2}}+2\frac{\partial_{r}\fEA}{\fEA}\frac{\partial_{r}\fEB}{\fEB}-\frac{(\partial_{r}\fEB)^{2}}{\fEB^{2}}-2\frac{\partial_{r}^{2}\fEB}{\fEB}\bigg) & =\fKappa\frho.
\end{align}
The vector constraints (\ref{eq:g-cc-vector})\textendash (\ref{eq:f-cc-vector})
are, 
\begin{align}
2\bigg[(\gK_{1}-\gK_{2})\frac{\partial_{r}\gEB}{\gEB}-\partial_{r}\gK_{2}\bigg] & =\uisignK\gKappa\gjota_{r},\\
2\bigg[(\fK_{1}-\fK_{2})\frac{\partial_{r}\fEB}{\fEB}-\partial_{r}\fK_{2}\bigg] & =\uisignK\fKappa\fjota_{r}.\label{eq:ssym-eq-j}
\end{align}
The last two equations can be recombined using the identity $\sqrt{\gSp}\,\gjota_{r}^{\gblab}+\sqrt{\fSp}\,\fjota_{r}^{\fblab}=0$,
\begin{align}
\fKappa\fEA\gEB\big(\gK_{1}\partial_{r}\gEB-\gK_{2}\partial_{r}\gEB-\gEB\partial_{r}\gK_{2}-\gKappa\gjota_{r}^{\mathrm{m}}\big)+\qquad\qquad\nonumber \\
\gKappa\gEA\fEB\big(\fK_{1}\partial_{r}\fEB-\fK_{2}\partial_{r}\fEB-\fEB\partial_{r}\fK_{2}-\fKappa\fjota_{r}^{\mathrm{m}}\big) & =0.
\end{align}
\eSe The radial separation parameter can be determined from (\ref{eq:ssym-eq-j})
as,
\begin{equation}
\sLp=\usignK\frac{2}{\gKappa\fEA\gEB\left\langle \sER\right\rangle _{1}^{2}}\big(\gK_{1}\partial_{r}\gEB-\gK_{2}\partial_{r}\gEB-\gEB\partial_{r}\gK_{2}\signK\gEB\gjota_{r}^{\mathrm{m}}\big).\label{eq:ssym-eq-p}
\end{equation}
The projection (\ref{eq:bim-prop-constr}) of the bimetric conservation
law reads,
\begin{align}
\Cbim & =\fEA\Big(\fK_{1}\left\langle \sER\right\rangle _{1}^{2}+2\fK_{2}\sER\left\langle \sER\right\rangle _{2}^{1}\Big)+2\gEA\fK_{2}\sLt\sER\left\langle \sER\right\rangle _{1}^{1}\nonumber \\
 & \qquad-\gEA\Big(\gK_{1}\left\langle \sER\right\rangle _{1}^{2}+2\gK_{2}\left\langle \sER\right\rangle _{1}^{1}\Big)-2\fEA\gK_{2}\sLt\left\langle \sER\right\rangle _{2}^{1}\nonumber \\
 & \qquad\isignK2\sLp\bigg(\left\langle \sER\right\rangle _{1}^{1}\frac{\gEA}{\fEA}\frac{\partial_{r}\fEB}{\gEB}+\left\langle \sER\right\rangle _{2}^{1}\frac{\fEA}{\gEA}\frac{\partial_{r}\gEB}{\gEB}\bigg)\isignK\sLtinv\left\langle \sER\right\rangle _{1}^{2}\partial_{r}\sLp=0,\label{eq:ssym-cc2}
\end{align}
where $\sLp$ can be eliminated using (\ref{eq:ssym-eq-p}). 

The evolution equations for the spatial metrics are,\bSe\label{eq:ssym-evol-1}
\begin{alignat}{2}
\partial_{t}\gEA & =\usignK\gAlpha\gEA\gK_{1}+\partial_{r}(\hShift\gEA+\gAlpha\sLv), & \qquad\partial_{t}\gEB & =\usignK\gAlpha\gEB\gK_{2}+\big(\hShift+\gAlpha\gEA^{-1}\sLv\big)\partial_{r}\gEB,\\
\partial_{t}\fEA & =\usignK\fAlpha\fEA\fK_{1}+\partial_{r}(\hShift\fEA-\fAlpha\sLv), & \partial_{t}\fEB & =\usignK\fAlpha\fEB\fK_{2}+\big(\hShift-\fAlpha\fEA^{-1}\sLv\big)\partial_{r}\fEB.
\end{alignat}
\eSe The evolution equations for the extrinsic curvatures are,\bSe\label{eq:ssym-evol-2}
\begin{align}
\partial_{t}\gK_{1} & =\big(\hShift+\gAlpha\gEA^{-1}\sLv\big)\partial_{r}\gK_{1}\isignK\gAlpha\gK_{1}\big(\gK_{1}+2\gK_{2}\big)\signK\gAlpha\gKappa\Big\{\,\gJota_{1}-\frac{1}{2}(\gJota-\grho)\,\Big\}\nonumber \\
 & \qquad\isignK\bigg(\frac{\partial_{r}\gAlpha}{\gEA^{2}}\frac{\partial_{r}\gEA}{\gEA}-\frac{\partial_{r}^{2}\gAlpha}{\gEA^{2}}+2\frac{\gAlpha}{\gEA^{2}}\frac{\partial_{r}\gEA}{\gEA}\frac{\partial_{r}\gEB}{\gEB}-2\frac{\gAlpha}{\gEA^{2}}\frac{\partial_{r}^{2}\gEB}{\gEB}\bigg),\\
\partial_{t}\fK_{1} & =\big(\hShift-\fAlpha\fEA^{-1}\sLv\big)\partial_{r}\fK_{1}\isignK\fAlpha\fK_{1}\big(\fK_{1}+2\fK_{2}\big)\signK\fAlpha\fKappa\Big\{\,\fJota_{1}-\frac{1}{2}(\fJota-\frho)\,\Big\}\nonumber \\
 & \qquad\isignK\bigg(\frac{\partial_{r}\fAlpha}{\fEA^{2}}\frac{\partial_{r}\fEA}{\fEA}-\frac{\partial_{r}^{2}\fAlpha}{\fEA^{2}}+2\frac{\fAlpha}{\fEA^{2}}\frac{\partial_{r}\fEA}{\fEA}\frac{\partial_{r}\fEB}{\fEB}-2\frac{\fAlpha}{\fEA^{2}}\frac{\partial_{r}^{2}\fEB}{\fEB}\bigg),\\
\partial_{t}\gK_{2} & =\big(\hShift+\gAlpha\gEA^{-1}\sLv\big)\partial_{r}\gK_{2}\isignK\gAlpha\gK_{2}\big(\gK_{1}+2\gK_{2}\big)\signK\gAlpha\gKappa\Big\{\,\gJota_{2}-\frac{1}{2}(\gJota-\grho)\,\Big\}\nonumber \\
 & \qquad\isignK\bigg(\frac{\gAlpha}{\gEB^{2}}-\frac{\partial_{r}\gAlpha}{\gEA^{2}}\frac{\partial_{r}\gEB}{\gEB}+\frac{\gAlpha}{\gEA^{2}}\frac{\partial_{r}\gEA}{\gEA}\frac{\partial_{r}\gEB}{\gEB}-\frac{\gAlpha}{\gEA^{2}}\frac{(\partial_{r}\gEB)^{2}}{\gEB^{2}}-\frac{\gAlpha}{\gEA^{2}}\frac{\partial_{r}^{2}\gEB}{\gEB}\bigg),\\
\partial_{t}\fK_{2} & =\big(\hShift-\fAlpha\fEA^{-1}\sLv\big)\partial_{r}\fK_{2}\isignK\fAlpha\fK_{2}\big(\fK_{1}+2\fK_{2}\big)\signK\fAlpha\fKappa\Big\{\,\fJota_{2}-\frac{1}{2}(\fJota-\frho)\,\Big\}\nonumber \\
 & \qquad\isignK\bigg(\frac{\fAlpha}{\fEB^{2}}-\frac{\partial_{r}\fAlpha}{\fEA^{2}}\frac{\partial_{r}\fEB}{\fEB}+\frac{\fAlpha}{\fEA^{2}}\frac{\partial_{r}\fEA}{\fEA}\frac{\partial_{r}\fEB}{\fEB}-\frac{\fAlpha}{\fEA^{2}}\frac{(\partial_{r}\fEB)^{2}}{\fEB^{2}}-\frac{\fAlpha}{\fEA^{2}}\frac{\partial_{r}^{2}\fEB}{\fEB}\bigg).
\end{align}
\eSe

\subsection{Bimetric conservation law revisited}

\label{sec:cons-law}

To determine the ratio of lapses we need to evaluate $\partial_{t}\Cbim$
from (\ref{eq:ssym-cc2}). Hence, in addition to the evolution of
the phase space variables (\ref{eq:ssym-evol-1}) and (\ref{eq:ssym-evol-2}),
we need to know the time derivative of the separation parameter, $\partial_{t}\sLp$.
In the following, we establish $\partial_{t}\sLp$ for the case of
spherical symmetry. We have not succeeded in calculating $\partial_{t}\sLp$
in the most general case. For the benefit of the reader, we have supplemented
the paper with one of the computations in appendix \ref{app:dtp}.

We begin by writing $\gCD_{\mu}\tud{\gVse}{\mu}{\nu}=0$ using the
ansatz (\ref{eq:ssym-gf}) and then projecting it. More specifically,
let $\gSector n^{\mu}=(\gAlpha^{-1},\gBeta\gAlpha^{-1},0,0)$ be the
unit normal on the spatial hypersurfaces. The projection along the
unit normal $\gSector n^{\nu}\gCD_{\mu}\tud{\gVse}{\mu}{\nu}=0$ yields,
\begin{align}
\left\langle \sER\right\rangle _{1}^{2}\partial_{t}\sLp & =\sLt\left\langle \sER\right\rangle _{1}^{2}\bigg(\frac{\partial_{r}\fAlpha}{\fEA}-\frac{\partial_{r}\gAlpha}{\gEA}\bigg)+\left\langle \sER\right\rangle _{1}^{2}\hShift\partial_{r}\sLp\nonumber \\
 & \qquad+\gAlpha\left\langle \sER\right\rangle _{1}^{1}\frac{2}{\gEB}\bigg(\frac{\partial_{r}\fEB}{\fEA}-\sLt\frac{\partial_{r}\gEB}{\gEA}\bigg)+\gAlpha\sLp\bigg(\gK_{1}\left\langle \sER\right\rangle _{1}^{2}+2\gK_{2}\left\langle \sER\right\rangle _{1}^{1}\bigg)\nonumber \\
 & \qquad+\fAlpha\left\langle \sER\right\rangle _{2}^{1}\frac{2}{\gEB}\bigg(\sLt\frac{\partial_{r}\fEB}{\fEA}-\frac{\partial_{r}\gEB}{\gEA}\bigg)+\fAlpha\sLp\bigg(-\frac{\left\langle \sER\right\rangle _{1}^{2}\partial_{r}\sLp}{\fEA\sLt}\bigg)\nonumber \\
 & \qquad+\fAlpha\left\langle \sER\right\rangle _{2}^{1}\frac{2}{\gEB}\frac{\partial_{r}\gEB}{\gEA}+\fAlpha\frac{2\gEA\left\langle \sER\right\rangle _{1}^{1}\partial_{r}\fEB}{\fEA^{2}\gEB}\nonumber \\
 & \qquad+\frac{\fAlpha}{\sLp}\bigg[-\frac{2\fK_{2}\gEA\sLt\left\langle \sER\right\rangle _{0}^{1}}{\fEA}+(-\frac{2\gEA\gK_{2}}{\fEA}-2\fK_{2}\sLt^{2})\left\langle \sER\right\rangle _{1}^{1}-2\gK_{2}\sLt\left\langle \sER\right\rangle _{2}^{1}\nonumber \\
 & \qquad\qquad+\frac{2\fK_{2}\gEA\sLt\left\langle \sER\right\rangle _{0}^{2}}{\fEA}+\left\langle \sER\right\rangle _{1}^{2}\bigl(-\frac{\gEA\gK_{1}}{\fEA}+(\fK_{1}+2\fK_{2})\sLt^{2}+\frac{\sLt\partial_{r}\sLp}{\fEA}\bigr)\bigg].\label{eq:dtp-rho}
\end{align}
 The same expression can also be obtained from the tangential projection,
\begin{equation}
(\tud{\delta}{\lambda}{\nu}+\gSector n^{\lambda}\gSector n_{\nu})\gCD_{\mu}\tud{\gVse}{\mu}{\lambda}=0.
\end{equation}
In this case,
\begin{align}
\left\langle \sER\right\rangle _{1}^{2}\partial_{t}\sLp & =\sLt\left\langle \sER\right\rangle _{1}^{2}\bigg(\frac{\partial_{r}\fAlpha}{\fEA}-\frac{\partial_{r}\gAlpha}{\gEA}\bigg)+\left\langle \sER\right\rangle _{1}^{2}\hShift\partial_{r}\sLp\nonumber \\
 & \qquad+\gAlpha\left\langle \sER\right\rangle _{1}^{1}\frac{2}{\gEB}\bigg(\frac{\partial_{r}\fEB}{\fEA}-\sLt\frac{\partial_{r}\gEB}{\gEA}\bigg)+\gAlpha\sLp\,\bigg(\gK_{1}\left\langle \sER\right\rangle _{1}^{2}+2\gK_{2}\left\langle \sER\right\rangle _{1}^{1}\bigg)\nonumber \\
 & \qquad+\fAlpha\left\langle \sER\right\rangle _{2}^{1}\frac{2}{\gEB}\bigg(\sLt\frac{\partial_{r}\fEB}{\fEA}-\frac{\partial_{r}\gEB}{\gEA}\bigg)+\fAlpha\sLp\,\bigg((\fK_{1}+2\fK_{2})\left\langle \sER\right\rangle _{1}^{2}-2\fK_{2}\left\langle \sER\right\rangle _{1}^{1}\bigg).\label{eq:dtp-j}
\end{align}
Now, the two evolution equations of $\sLp$ must be equal at all times,
that is, their difference must vanish identically. Subtracting (\ref{eq:dtp-j})
from (\ref{eq:dtp-rho}) and multiplying by $\sLp\fEA/\fAlpha$ gives,
\begin{align}
0 & =\left\langle \sER\right\rangle _{1}^{2}\frac{\partial_{r}\sLp}{\sLt}+2\sLp\bigg(\left\langle \sER\right\rangle _{1}^{1}\frac{\gEA}{\fEA}\frac{\partial_{r}\fEB}{\gEB}+\left\langle \sER\right\rangle _{2}^{1}\frac{\fEA}{\gEB}\frac{\partial_{r}\gEB}{\gEA}\bigg)\nonumber \\
 & \qquad+2\sLt\left(\fK_{2}\gEA\sER\left\langle \sER\right\rangle _{1}^{2}-\fEA\gK_{2}\left\langle \sER\right\rangle _{2}^{1}\right)\nonumber \\
 & \qquad+\fEA\left(\fK_{1}\left\langle \sER\right\rangle _{1}^{2}+2\fK_{2}\sER\left\langle \sER\right\rangle _{2}^{1}\right)-\gEA\gK_{1}\left\langle \sER\right\rangle _{1}^{2}-2\gEA\gK_{2}\left\langle \sER\right\rangle _{1}^{1},
\end{align}
which is the bimetric conservation law. We again obtain (\ref{eq:ssym-cc2}),
which is consistent with (\ref{eq:bim-prop-constr}) for the case
of spherical symmetry. This is not surprising as the above procedure
reproduces the steps from Lemma 1 in \cite{Kocic:2018ddp}.

\subsection{The ratio of lapses}

\label{sec:ratio}

The relation between the two lapses is a consequence of the requirement
that the projected bimetric conservation law (\ref{eq:ssym-cc2})
stays preserved in time, that is, $\partial_{t}\Cbim=0$. To evaluate
$\partial_{t}\Cbim$, we need to know the evolution equations for
all the fields, in particular $\sLp$. The evolution of $\sLp$ is
given by (\ref{eq:dtp-j}). The time derivative of (\ref{eq:ssym-cc2})
gives,
\begin{align}
0 & =F_{1}\partial_{t}\fEA+F_{2}\partial_{t}\fEB+F_{3}\partial_{t}\fK_{1}+F_{4}\partial_{t}\fK_{2}\nonumber \\
 & \qquad+\,F_{5}\partial_{t}\gEA+F_{6}\partial_{t}\gEB+F_{7}\partial_{t}\gK_{1}+F_{8}\partial_{t}\gK_{2}\nonumber \\
 & \qquad+\,F_{9}\partial_{tr}\fEB+F_{10}\partial_{tr}\gEB+F_{11}.
\end{align}
The coefficients $F_{i}$ are lengthy and given in the ancillary Mathematica
notebook. After replacing the time derivatives with the respective
evolution equations (\ref{eq:ssym-evol-1}), (\ref{eq:ssym-evol-2}),
and (\ref{eq:dtp-j}), one gets,
\begin{equation}
0=G_{0}+G_{1}\gAlpha+G_{2}\fAlpha+G_{3}\partial_{r}\gAlpha+G_{4}\partial_{r}\fAlpha+G_{5}\partial_{r}^{2}\gAlpha+G_{6}\partial_{r}^{2}\fAlpha.
\end{equation}
 The coefficients $G_{i}$ are even more complicated than $F_{i}$.
However, it is straightforward to show $G_{5}=G_{6}=0$, and to conclude
that $G_{3}=G_{4}=0$ (again, the details are given in the ancillary
Mathematica notebook). Furthermore, the first term can be rewritten,
\begin{equation}
G_{0}=\partial_{r}\left(\hShift\Cbim\right)\approx0,
\end{equation}
since $\Cbim=0$. Therefore, we are left with the weak equality,
\begin{equation}
0\approx\partial_{t}\Cbim\approx G_{1}\gAlpha+G_{2}\fAlpha.\label{eq:lapseeq1}
\end{equation}
We can rewrite (\ref{eq:lapseeq1}) as,
\begin{equation}
0\approx\frac{\gW}{W_{c}}\gAlpha+\frac{\fW}{W_{c}}\fAlpha,\label{lapseeq2}
\end{equation}
where,
\begin{equation}
G_{1}=\frac{\gW}{W_{c}}=\frac{\gCW}{\sqrt{\gSp}},\qquad G_{2}=\frac{\fW}{W_{c}}=\frac{\fCW}{\sqrt{\gSp}}.
\end{equation}
Equation (\ref{lapseeq2}) is (\ref{lapseeq}) in the spherically
symmetric case. 

The variables $\gW$ and $\fW$ depend on the spatial metrics, the
extrinsic curvatures, their derivatives, and on the contributions
from the matter stress\textendash energy tensors (hence, coupling
matter to any of the metrics automatically affects the lapse function
of the other metric). However, the second derivatives of the metric
components can be eliminated using the scalar constraints, the derivatives
of the extrinsic curvatures can be solved using the vector constraints,
and the derivative $\partial_{r}\sLp$ can be eliminated using $\Cbim=0$.
This gives the final expression,
\begin{align}
W_{c} & =2\fEA^{2}\gEA^{2}\gEB^{2}\sER^{2}\sLt^{3}\left\langle \sER\right\rangle _{1}^{2},\\[1ex]
\gW & =\sLt^{2}\bigl[c_{1}+\sLt(c_{2}+\sLt c_{3})+\sLp^{2}c_{4}\bigl]\nonumber \\
 & \quad+\fKappa\sLt^{2}(c_{5}+\sLt c_{6})+\gKappa\sLt^{2}\bigl[c_{7}+\sLt^{2}c_{9}+\sLt(c_{8}+\sLp c_{10})+\sLp^{2}c_{11}\bigl]\nonumber \\
 & \quad+\sLt\sLp\bigl[c_{12}+\sLt(c_{13}+\sLt c_{14})+\sLp^{2}c_{15}\bigl]\partial_{r}\fEB+\sLt^{2}(c_{16}+\sLt c_{17})(\partial_{r}\fEB)^{2}\nonumber \\
 & \quad+\sLt^{2}\sLp(c_{18}+\sLt c_{19})\partial_{r}\gEB+\sLt^{3}(c_{20}+\sLt c_{21})\partial_{r}\fEB\partial_{r}\gEB\nonumber \\
 & \quad+\sLt^{2}\bigl[\sLt(c_{22}+\sLt c_{23})+\sLp^{2}c_{24}\bigl](\partial_{r}\gEB)^{2},\\[1ex]
\fW & =\sLt^{2}\bigl[d_{1}+\sLt(d_{2}+\sLt d_{3})+\sLp^{2}d_{4}\bigl]\nonumber \\
 & \quad+\fKappa\sLt^{2}\bigl[d_{5}+\sLt^{2}d_{7}+\sLt(d_{6}+\sLp d_{8})+\sLp^{2}d_{9}\bigl]+\gKappa\sLt^{2}(d_{10}+\sLt d_{11})\nonumber \\
 & \quad+\sLt^{2}\sLp(d_{12}+\sLt d_{13})\partial_{r}\fEB+\sLt^{2}\bigl[\sLt(d_{14}+\sLt d_{15})+\sLp^{2}d_{16}\bigl](\partial_{r}\fEB)^{2}\nonumber \\
 & \quad+\sLt\sLp\bigl[d_{17}+\sLt(d_{18}+\sLt d_{19})+\sLp^{2}d_{20}\bigl]\partial_{r}\gEB+\sLt^{3}(d_{21}+\sLt d_{22})\partial_{r}\fEB\partial_{r}\gEB\nonumber \\
 & \quad+\sLt^{2}(d_{23}+\sLt d_{24})(\partial_{r}\gEB)^{2}.
\end{align}
The coefficients $c_{i}$ and $d_{i}$ are given in appendix \ref{sec:cd-coef}.
They do not contain any derivatives of the fields. As expected, the
coefficients are symmetric with respect to the duality $\gMet\leftrightarrow\fMet$,
$\beta_{4-n}\leftrightarrow\beta_{n}$, $\sLp\leftrightarrow-\sLp$,
and $\sgn\leftrightarrow-\sfn$. Under this exchange, we have $\gW\leftrightarrow-\fW$.

\section{Discussion}

\label{sec:discussion}

In order to set up the initial data in HR bimetric relativity, it
is necessary to solve the constraints (\ref{eq:g-cc}) and (\ref{eq:bim-prop-constr}).
In addition, to find the development, one must also ensure that $\partial_{t}\Ctwo=0$.
As shown in \cite{Hassan:2018mbl,Alexandrov:2012yv}, this condition
tells us that the lapse functions of the two metrics are not independent,
but that their ratio, $\mSector W$, is a function of the dynamical
variables (the spatial metric functions and their conjugate momenta
or alternatively the extrinsic curvatures) and their spatial derivatives.
Knowing $\mSector W$ makes it possible to evolve the initial data.

To compute $\mSector W$, the equations of motion for all the fields
are required. In particular, we need $\partial_{t}\sLp$, that is,
the time derivative of the Lorentz vector parametrizing the difference
between the shifts of the two metrics. Computing $\partial_{t}\sLp$
in the most general case is not straightforward. We have tried several
approaches, but neither of them worked out. For the benefit of the
reader, one of the computations is supplemented in appendix \ref{app:dtp}.
Nevertheless, we have calculated $\partial_{t}\sLp$, and hence $\mSector W$,
for the case of spherical symmetry (see section \ref{sec:main}).

Since one lapse is given in terms of the other lapse,
\begin{equation}
\gAlpha=\mSector W\fLapse=-\frac{\fW}{\gW}\fAlpha,\qquad\fAlpha=\mSector W^{-1}\gAlpha=-\frac{\gW}{\fW}\gAlpha,\label{eq:gauge1}
\end{equation}
the gauge condition can be imposed on either one of them, and the
other one is determined. It is also possible to relate the lapse $\hLapse$
of the geometric mean metric $\hMet=\gMet\big(\gMet^{-1}\fMet\big)^{1/2}$
to $\gAlpha$ and $\fAlpha$ through \cite{Hassan:2017ugh,Kocic:2018ddp},
\begin{equation}
\hLapse^{2}\sLt=\gAlpha\fAlpha,
\end{equation}
 and express $\gAlpha$ and $\fAlpha$ in terms of $\hLapse$,
\begin{equation}
\gAlpha=\hLapse\sqrt{-\sLt\frac{\fW}{\gW}}=\hLapse\sqrt{\sLt\mSector W},\qquad\fAlpha=\hLapse\sqrt{-\sLt\frac{\gW}{\fW}}=\hLapse\sqrt{\frac{\sLt}{\mSector W}}.\label{eq:gauge2}
\end{equation}
 This gives several possible choices when adapting the geometry of
foliations. The gauge choices with respect to $\hMet$ are the subject
of \cite{Torsello:meang}.

Finally, cases where $\fW$ or $\gW$, or both, vanish or diverge
for some values of the fields must be dealt with using a suitable
choice for the slicing. 

%\acknowledgments\vspace{-1.8ex}
\subsubsection*{Acknowledgments}

We are grateful to Fawad Hassan, Edvard M\"{o}rtsell, and Marcus
H\"{o}g\r{a}s for the valuable discussions. We also wish to thank
them and Angnis Schmidt-May for a careful reading of the manuscript. 

\begin{comment}
Finally, an interesting physical interpretation of the ratio of lapses
comes from the fact that a lapse function measures the proper time
of an Eulerian observer whose worldlines are orthogonal to the spatial
hypersurfaces relative to the metric. In this view, the ratio of lapses
represents the ratio of the proper times of the two Eulerian observers
for $\gMet$ and $\fMet$,  
\begin{equation}
\frac{\dd\tau_{\gMet}}{\dd\tau_{\fMet}}=\frac{\gAlpha\,\dd t}{\fLapse\,\dd t}=\mSector W.
\end{equation}
\end{comment}

%%%%%%%%%%%%%%%%%%%%%%%%%%%%%%%%%%%%%%%%%%%%%%%%%%%%%%%%%%%%%%%%%%%%%%%%%%%%%
% Appendices
%%%%%%%%%%%%%%%%%%%%%%%%%%%%%%%%%%%%%%%%%%%%%%%%%%%%%%%%%%%%%%%%%%%%%%%%%%%%%

\clearpage
\emitAppendix

\begin{comment}
``Lasciate ogni speranza, o voi che entrate.''
\end{comment}

\section{Explicit computation of Poisson brackets}

\label{app:poiss}

In order to keep the derivation in subsection \ref{cratio} brief,
a number of intermediate steps were skipped. Those steps are shown
explicitly here. Equation \eqref{C2Ri} can be derived in the following
way, using the Jacobi identity as well as equation \eqref{CLRi},
\eqref{CNRi}, and \eqref{CLCN},
\begin{align}
\{\Ctwo(x),\gfCvec_{i}(y)\} & =\int\dd^{3}z\{\{\CL(x),\CN(z)\},\gfCvec_{i}(y)\}\nonumber \\
 & =\int\dd^{3}z\left(\{\CL(x),\{\CN(z),\gfCvec_{i}(y)\}\}-\{\CN(z),\{\CL(x),\gfCvec_{i}(y)\}\}\right)\nonumber \\
 & =\int\dd^{3}z\left(-\{\CL(x),\CN(y)\}\frac{\partial}{\partial z^{i}}\delta^{3}(z-y)+\{\CN(z),\CL(y)\}\frac{\partial}{\partial x^{i}}\delta^{3}(x-y)\right)\nonumber \\
 & =\int\dd^{3}z\left(-\Ctwo(x)\delta^{3}(x-y)\frac{\partial}{\partial z^{i}}\delta^{3}(z-y)-\Ctwo(y)\delta^{3}(y-z)\frac{\partial}{\partial x^{i}}\delta^{3}(x-y)\right)\nonumber \\
 & =-\Ctwo(y)\frac{\partial}{\partial x^{i}}\delta^{3}(x-y),\label{C2Ri-1}
\end{align}
where, in the last step, the first term vanishes due to integration
by parts.

Equation \eqref{eq:C2CL} states that,\vspace{-0.1ex}
\begin{align}
\{\Ctwo(x),\CL(y)\} & =\int\dd^{3}z\left(\{\CL(x),\{\CN(z),\CL(y)\}\}-\{\CN(z),\{\CL(x),\CL(y)\}\}\right).
\end{align}
The last term of this can be rewritten,\vspace{-0.1ex}
\begin{align}
 & \int\dd^{3}z\{\CN(z),\{\CL(x),\CL(y)\}\}\nonumber \\
= & -\int\dd^{3}z\left\{ \CN(z),\fSp^{ij}(x)\gfCvec_{j}(x)\frac{\partial}{\partial x^{i}}\delta^{3}(x-y)-\fSp^{ij}(y)\gfCvec_{j}(y)\frac{\partial}{\partial y^{i}}\delta^{3}(x-y)\right\} \nonumber \\
= & -\int\dd^{3}z\left(\fSp^{ij}(x)\{\CN(z),\gfCvec_{j}(x)\}+\fSp^{ij}(y)\{\CN(z),\gfCvec_{j}(x)\}\right)\frac{\partial}{\partial x^{i}}\delta^{3}(x-y)\nonumber \\
= & \int\dd^{3}z\left(\fSp^{ij}(x)\CN(x)\frac{\partial}{\partial z^{i}}\delta^{3}(z-x)+\fSp^{ij}(y)\CN(y)\frac{\partial}{\partial z^{i}}\delta^{3}(z-y)\right)\frac{\partial}{\partial x^{i}}\delta^{3}(x-y),
\end{align}
where we have made use of \eqref{CLCL} and \eqref{CNRi}, as well
as the fact that $\CN$ does not depend on $\fPi^{ij}$. Using integration
by parts, it follows that this vanishes strongly. Hence, equation
\eqref{eq:C2CL} reduces to the first equality in \eqref{C2CLsimp}.

In a similar way, consider equation \eqref{eq:C2CN}, 
\begin{align}
\{\Ctwo(x),\CN(y)\} & =\int\dd^{3}z\left(\{\CL(z),\{\CN(x),\CN(y)\}\}-\{\CN(x),\{\CL(z),\CN(y)\}\}\right).
\end{align}
The first term can be rewritten,\vspace{-0.1ex}
\begin{align}
 & \int\dd^{3}z\{\CL(z),\{\CN(x),\CN(y)\}\}\nonumber \\
=- & \int\dd^{3}z\left\{ \CL(z),\CN(x)\sgn^{i}(x)\frac{\partial}{\partial x^{i}}\delta^{3}(x-y)-\CN(y)\sgn^{i}(y)\frac{\partial}{\partial y^{i}}\delta^{3}(x-y)\right\} \nonumber \\
=- & \int\dd^{3}z\left(\{\CL(z),\CN(x)\sgn^{i}(x)\}+\{\CL(z),\CN(y)\sgn^{i}(y)\}\right)\frac{\partial}{\partial x^{i}}\delta^{3}(x-y)\nonumber \\
=- & \int\dd^{3}z\Big[\{\CL(z),\CN(x)\}\sgn^{i}(x)+\{\CL(z),\sgn^{i}(x)\}\CN(x)\nonumber \\
 & \qquad\quad+\{\CL(z),\CN(y)\}\sgn^{i}(y)+\{\CL(z),\sgn^{i}(y)\}\CN(y)\Big]\frac{\partial}{\partial x^{i}}\delta^{3}(x-y),
\end{align}
\clearpage\noindent where \eqref{CNCN} has been used. Using \eqref{CLCN},
it follows that every term in the final expression is proportional
to either $\CN$ or $\Ctwo$, meaning that this entire expression
vanishes weakly. Thus, equation \eqref{eq:C2CN} reduces to the first
equality in \eqref{C2CNsimp}.

\section{3+1 bimetric variables}

\label{app:bim-vars}

The primary variables are the lapses $\gLapse$, $\fLapse$, the spatial
vielbeins $\gE$, $\fE$, the overall mean-shift vector $\hShiftVec$,
and the Lorentz vector $\sLp$ that defines the relative shift between
the metrics. The separation parameter $\sLp$ is in the boost parameter
of the Lorentz transformation that contains the spatial part $\sLs$
and the Lorentz factor $\sLt$ where $\sLp=\sLs\sLv=\sLt\sLv$ and,
\begin{align}
\sLt & \coloneqq\big(1+\sLp^{{\scriptscriptstyle \tr}}\sEta\sLp\big)^{1/2}=\big(1-\sLv^{{\scriptscriptstyle \tr}}\sEta\sLv\big)^{-1/2},\\
\sLs & \coloneqq\big(\sI+\sLp\sLp^{\tr}\sEta\big)^{1/2}=\big(\sI-\sLv\sLv^{\tr}\sEta\big)^{-1/2},
\end{align}
where $\sI$ denotes the spatial identity and $\sEta$ is the spatial
part of the Minkowski metric. 

The variables derived from $\gLapse$, $\gE$, $\fLapse$, $\fE$,
$\sLp$, and $\hShiftVec$ are (for details see \cite{Kocic:2018ddp}),
\begin{align}
\gSp & \coloneqq\gE^{{\scriptscriptstyle \tr}}\sEta\gE, & \fSp & \coloneqq\fE^{{\scriptscriptstyle \tr}}\sEta\fE,\\
\sgn & \coloneqq\gE^{-1}\sLv, & \sfn & \coloneqq\fE^{-1}\sLv,\\
\gShiftVec & \coloneqq\hShiftVec+\gLapse\sgn, & \fShiftVec & \coloneqq\hShiftVec-\fLapse\sfn,\\
\sgQ & \coloneqq\gE^{-1}\sLs^{2}\gE, & \sfQ & \coloneqq\fE^{-1}\sLs^{2}\fE,\\
\sgD & \coloneqq\fE^{-1}\sLs^{-1}\gE, & \sfD & \coloneqq\gE^{-1}\sLs^{-1}\fE,\\
\sgB & \coloneqq\sgD^{-1}=\gE^{-1}\sLs\fE=\sfD\sfQ=\sgQ, & \sfB & \coloneqq\sfD^{-1}=\fE^{-1}\sLs\gE=\sgD\sgQ=\sfQ,\\
\sgV & \coloneqq\usignV\betaSum\,e_{n}(\sfD), & \sfV & \coloneqq\usignV\betaSum\,\sLtinv e_{n-1}(\sgB),\\
\sgU & \coloneqq\usignV\betaSum\,\sLtinv Y_{n-1}(\sgB), & \sfU & \coloneqq\usignV\betaSum\,\sfD\,Y_{n-1}(\sfD),\\
\sgQU & \coloneqq\usignV\betaSum\,\sgB\,Y_{n-1}(\sfD), & \sfQU & \coloneqq\usignV\betaSum\,\sLtinv\sfQ\,Y_{n-1}(\sgB).
\end{align}
Here $e_{n}$ denotes the elementary symmetric polynomials while $Y_{n}$
stands for their derivatives,
\begin{equation}
Y_{n}(S)\coloneqq\sum_{k=0}^{n}(-1)^{n+k}e_{k}(S)\,S^{n-k}=\frac{\partial e_{n+1}(S)}{\partial S^{\tr}}.
\end{equation}

\paragraph*{Restoring indices.}

We use the convention where $i,j,...$ denote the spatial world indices,
and $a,b,...$ denote the spatial Lorentz indices. Note that $\gLapse$,
$\fLapse$, $\sgV$, $\sfV$, and $\sLt$ are scalars.
\begin{itemize}[itemsep=-0.6ex]
\item In the world frame we have $\gSp_{ij}$, $\gShiftVec^{i}$, $\sgn^{i}$,
$\fSp_{ij}$, $\fShiftVec^{i}$, $\sfn^{i}$, and $\hShiftVec^{i}$. 
\item In the Lorentz frame we have $\sLp^{a}$, $\sLv^{a}$, $\tud{\sLs}ab$,
$\sEta_{ab}=\delta_{ab}$, and $\tud{\sI}ab=\delta_{b}^{a}$. 
\item In the mixed frame we have $\tud{\gE}ai$ and $\tud{\fE}ai$. Note
that, for instance, $\gE^{{\scriptscriptstyle \tr}}$ and $\gE^{-1}$
become $(\gE^{{\scriptscriptstyle \tr}}\tdu )ia$ and $(\gE^{-1}\tud )ia$,
respectively. 
\item All other symbols represent the spatial operators; for example, $\sgD$
becomes $\tud{\sgD}ij$. The compound symbol $\sgQU$ denotes $\sgQ\sfU$;
hence, $\tud{\sgQU}ij=\tud{\sgQ}ik\tud{\sfU}kj$.
\end{itemize}

\clearpage

\section{Spherically symmetric variables}

\label{app:ssym-vars}

Here we reduce the 3+1 variables to the spherically symmetric case,
where $\gBeta=\hShift+\gAlpha\gEA^{-1}\sLp\sLtinv$ and $\fBeta=\hShift-\fAlpha\fEA^{-1}\sLp\sLtinv$.
Note that $\sLv=\sLp\sLtinv$ and $\sLt^{2}=1+\sLp^{2}$. 

The variables are,
\begin{alignat}{2}
\sgn^{r} & =\gEA^{-1}\sLp\sLtinv, & \sfn^{r} & =\fEA^{-1}\sLp\sLtinv,\\
\tud{\sgQ}rr & =\sLt^{2}, & \tud{\sfQ}rr & =\sLt^{2},\\
\tud{\sgQ}{\theta}{\theta} & =1, & \tud{\sfQ}{\theta}{\theta} & =1,\\
\tud{\sgD}rr & =\sLtinv\gEA\fEA^{-1}, & \tud{\sfD}rr & =\sLtinv\fEA\gEA^{-1},\\
\tud{\sgD}{\theta}{\theta} & =\gEB\fEB^{-1}=\sER^{-1}, & \tud{\sfD}{\theta}{\theta} & =\fEB\gEB^{-1}=\sER,\\
\tud{\sgB}rr & =\sLt\fEA\gEA^{-1}, & \tud{\sfB}rr & =\sLt\gEA\fEA^{-1},\\
\tud{\sgB}{\theta}{\theta} & =\fEB\gEB^{-1}=\sER, & \tud{\sfB}{\theta}{\theta} & =\gEB\fEB^{-1}=\sER^{-1},\\
\sgV & =\left\langle \sER\right\rangle _{0}^{2}+\sLtinv\fEA\gEA^{-1}\left\langle \sER\right\rangle _{1}^{2}, & \sfV & =\sLtinv\left\langle \sER\right\rangle _{1}^{2}+\fEA\gEA^{-1}\left\langle \sER\right\rangle _{2}^{2},\\
\tud{\sgU}rr & =\sLtinv\left\langle \sER\right\rangle _{1}^{2}, & \tud{\sfU}rr & =\sLtinv\fEA\gEA^{-1}\left\langle \sER\right\rangle _{1}^{2},\\
\tud{\sgU}{\theta}{\theta} & =\sLtinv\left\langle \sER\right\rangle _{1}^{1}+\fEA\gEA^{-1}\left\langle \sER\right\rangle _{2}^{1}, & \tud{\sfU}{\theta}{\theta} & =\sER\left\langle \sER\right\rangle _{1}^{1}+\sLtinv\fEA\gEA^{-1}\sER\left\langle \sER\right\rangle _{2}^{1},\\
\tud{\sgQU}rr & =\sLt\fEA\gEA^{-1}\left\langle \sER\right\rangle _{1}^{2}, & \tud{\sfQU}rr & =\sLt\left\langle \sER\right\rangle _{1}^{2},\\
\tud{\sgQU}{\theta}{\theta} & =\sER\left\langle \sER\right\rangle _{1}^{1}+\sLtinv\fEA\gEA^{-1}\sER\left\langle \sER\right\rangle _{2}^{1},\qquad & \tud{\sfQU}{\theta}{\theta} & =\sLtinv\left\langle \sER\right\rangle _{1}^{1}+\fEA\gEA^{-1}\left\langle \sER\right\rangle _{2}^{1},
\end{alignat}
where,
\begin{equation}
\left\langle \sER\right\rangle _{k}^{1}=\signV m^{4}\big(\beta_{(k)}+\beta_{(k+1)}\sER\big),\quad\left\langle \sER\right\rangle _{k}^{2}=\signV m^{4}\big(\beta_{(k)}+2\beta_{(k+1)}\sER+\beta_{(k+2)}\sER^{2}\big).
\end{equation}
All other components are zero. For any spatial operator $X$, we have
$\tud X{\theta}{\theta}=\tud X{\phi}{\phi}$. 

The nonzero components of the projections of the bimetric stress\textendash energy
tensor $V_{\gMet}$ are,\bSe\label{eq:ssym-g-se}\vspace{-1ex}
\begin{align}
\grho^{\gblab} & =-\Bigg[\left\langle \sER\right\rangle _{0}^{2}+\sLt\frac{\fEA}{\gEA}\left\langle \sER\right\rangle _{1}^{2}\Bigg],\qquad\gjota_{r}^{\gblab}=-\sLp\fEA\left\langle \sER\right\rangle _{1}^{2},\\
\gJota_{1}^{\gblab} & =\left\langle \sER\right\rangle _{0}^{2}+\Bigg[\frac{1}{\sLt}\bigg(\frac{\fAlpha}{\gAlpha}+\frac{\fEA}{\gEA}\bigg)-\sLt\frac{\fEA}{\gEA}\Bigg]\left\langle \sER\right\rangle _{1}^{2},\\
\gJota_{2}^{\gblab} & =\left\langle \sER\right\rangle _{0}^{1}+\frac{\fAlpha\fEA}{\gAlpha\gEA}\left\langle \sER\right\rangle _{1}^{2}+\frac{1}{\sLt}\bigg(\frac{\fAlpha}{\gAlpha}+\frac{\fEA}{\gEA}\bigg)\left\langle \sER\right\rangle _{1}^{1}.
\end{align}
\eSe Similarly for $V_{\fMet}$ we have,\bSe \label{eq:ssym-f-se}\vspace{-1ex}
\begin{align}
\frho^{\fblab} & =-\Bigg[\left\langle \sER\right\rangle _{2}^{2}+\sLt\frac{\gEA}{\fEA}\left\langle \sER\right\rangle _{1}^{2}\Bigg]\frac{1}{\sER^{2}},\qquad\fjota_{r}^{\fblab}=\sLp\gEA\left\langle \sER\right\rangle _{1}^{2}\frac{1}{\sER^{2}},\\
\fJota_{1}^{\fblab} & =\Bigg\{\left\langle \sER\right\rangle _{2}^{2}+\Bigg[\frac{1}{\sLt}\bigg(\frac{\gAlpha}{\fAlpha}+\frac{\gEA}{\fEA}\bigg)-\sLt\frac{\gEA}{\fEA}\Bigg]\left\langle \sER\right\rangle _{1}^{2}\Bigg\}\frac{1}{\sER^{2}},\\
\fJota_{2}^{\fblab} & =\Bigg\{\left\langle \sER\right\rangle _{3}^{1}+\frac{\gAlpha\gEA}{\fAlpha\fEA}\left\langle \sER\right\rangle _{1}^{1}+\frac{1}{\sLt}\bigg(\frac{\gAlpha}{\fAlpha}+\frac{\gEA}{\fEA}\bigg)\left\langle \sER\right\rangle _{2}^{1}\Bigg\}\frac{1}{\sER},
\end{align}
\eSe

\section{Coefficients in the lapse ratio}

\label{sec:cd-coef}

\begin{align*}
c_{1} & =2\fEA^{3}\gEA^{2}\sER\bigl[\fK_{2}^{2}\gEB^{2}\sER^{2}\left\langle \sER\right\rangle _{1}^{1}(4\left\langle \sER\right\rangle _{1}^{1}-3\left\langle \sER\right\rangle _{1}^{2})+\left\langle \sER\right\rangle _{1}^{1}\left\langle \sER\right\rangle _{1}^{2}-\sER\left\langle \sER\right\rangle _{2}^{1}\left\langle \sER\right\rangle _{1}^{2}\bigl],\\
c_{2} & =2\fEA^{2}\gEA^{2}\sER^{2}\Bigl(2\fEA\fK_{2}\gEB^{2}\gK_{2}\sER(4\left\langle \sER\right\rangle _{1}^{1}\left\langle \sER\right\rangle _{2}^{1}-\left\langle \sER\right\rangle _{2}^{0}\left\langle \sER\right\rangle _{1}^{2})\\
 & \quad+\gEA\Bigl\{-4\fK_{2}^{2}\gEB^{2}\sER^{2}(\left\langle \sER\right\rangle _{1}^{1})^{2}+\left\langle \sER\right\rangle _{1}^{2}\bigl[-2(1+2\gEB^{2}\gK_{2}^{2})\left\langle \sER\right\rangle _{1}^{1}+(1+\gEB^{2}\gK_{2}^{2})\left\langle \sER\right\rangle _{1}^{2}\bigl]\\
 & \quad+2\gEB^{2}\gK_{2}^{2}\sER(2\left\langle \sER\right\rangle _{1}^{1}\left\langle \sER\right\rangle _{2}^{1}-\left\langle \sER\right\rangle _{2}^{0}\left\langle \sER\right\rangle _{1}^{2})\Bigl\}\Bigr),\\
c_{3} & =2\fEA^{2}\gEA^{2}\gEB^{2}\gK_{2}\sER^{2}\Bigl\{\fEA\gK_{2}\bigl[4\sER(\left\langle \sER\right\rangle _{2}^{1})^{2}-2\sER\left\langle \sER\right\rangle _{3}^{0}\left\langle \sER\right\rangle _{1}^{2}-3\left\langle \sER\right\rangle _{2}^{1}\left\langle \sER\right\rangle _{1}^{2}\bigl]\\
 & \quad+2\fK_{2}\gEA\sER\bigl[2(\left\langle \sER\right\rangle _{1}^{1})^{2}+\sER\left\langle \sER\right\rangle _{2}^{0}\left\langle \sER\right\rangle _{1}^{2}+\left\langle \sER\right\rangle _{1}^{1}(-2\sER\left\langle \sER\right\rangle _{2}^{1}+\left\langle \sER\right\rangle _{1}^{2})\bigl]\Bigl\},\\
c_{4} & =2\fEA^{3}\gEA^{2}\gEB^{2}\gK_{2}^{2}\sER^{2}\left\langle \sER\right\rangle _{2}^{1}(-4\left\langle \sER\right\rangle _{1}^{1}+\left\langle \sER\right\rangle _{1}^{2}),\\
c_{5} & =\fEA^{3}\gEA^{2}\gEB^{2}\left\langle \sER\right\rangle _{1}^{2}\bigl[\left\langle \sER\right\rangle _{1}^{2}(2\sER\left\langle \sER\right\rangle _{2}^{1}+\left\langle \sER\right\rangle _{1}^{2})-2\left\langle \sER\right\rangle _{1}^{1}(\frhom\sER^{3}+\left\langle \sER\right\rangle _{1}^{2}-\sER\left\langle \sER\right\rangle _{2}^{2})\bigl],\\
c_{6} & =6\fEA^{2}\gEA^{3}\gEB^{2}\sER\left\langle \sER\right\rangle _{1}^{1}(\left\langle \sER\right\rangle _{1}^{2})^{2},\\
c_{7} & =\fEA^{3}\gEA^{2}\gEB^{2}\sER^{2}\left\langle \sER\right\rangle _{1}^{2}\bigl[\left\langle \sER\right\rangle _{1}^{2}(\left\langle \sER\right\rangle _{1}^{2}-4\left\langle \sER\right\rangle _{1}^{1})-2\left\langle \sER\right\rangle _{2}^{1}\left\langle \sER\right\rangle _{0}^{2}\bigl],\\
c_{8} & =2\fEA^{2}\gEA^{3}\gEB^{2}\sER^{2}\left\langle \sER\right\rangle _{1}^{2}\bigl[2\left\langle \sER\right\rangle _{1}^{1}(\gJotam 2-\left\langle \sER\right\rangle _{0}^{2})+(\gJotam 1-\grhom-\left\langle \sER\right\rangle _{0}^{1}+\left\langle \sER\right\rangle _{0}^{2})\left\langle \sER\right\rangle _{1}^{2}\bigl],\\
c_{9} & =2\fEA^{3}\gEA^{2}\gEB^{2}(2\gJotam 2-\grhom)\sER^{2}\left\langle \sER\right\rangle _{2}^{1}\left\langle \sER\right\rangle _{1}^{2},\\
c_{10} & =4\fEA^{3}\gEA\gEB^{2}\gjotam\sER^{2}\left\langle \sER\right\rangle _{2}^{1}\left\langle \sER\right\rangle _{1}^{2},\\
c_{11} & =-2\fEA^{3}\gEA^{2}\gEB^{2}\grhom\sER^{2}\left\langle \sER\right\rangle _{2}^{1}\left\langle \sER\right\rangle _{1}^{2},\\
c_{12} & =8\fEA\gEA^{3}\gEB\gK_{2}\sER^{2}(\left\langle \sER\right\rangle _{1}^{1})^{2},\\
c_{13} & =4\fEA^{2}\gEA^{2}\gEB\gK_{2}\sER^{2}(\left\langle \sER\right\rangle _{2}^{0}\left\langle \sER\right\rangle _{1}^{2}-4\left\langle \sER\right\rangle _{1}^{1}\left\langle \sER\right\rangle _{2}^{1}),\\
c_{14} & =4\fEA\gEA^{3}\gEB\gK_{2}\sER^{2}\bigl[\sER\left\langle \sER\right\rangle _{2}^{0}\left\langle \sER\right\rangle _{1}^{2}+\left\langle \sER\right\rangle _{1}^{1}(\left\langle \sER\right\rangle _{1}^{2}-2\sER\left\langle \sER\right\rangle _{2}^{1})\bigl],\\
c_{15} & =8\fEA\gEA^{3}\gEB\gK_{2}\sER^{2}(\left\langle \sER\right\rangle _{1}^{1})^{2},\\
c_{16} & =-2\fEA\gEA^{2}\sER\bigl[\left\langle \sER\right\rangle _{1}^{1}(4\sER\left\langle \sER\right\rangle _{2}^{1}+\left\langle \sER\right\rangle _{1}^{2})-2\sER\left\langle \sER\right\rangle _{2}^{0}\left\langle \sER\right\rangle _{1}^{2}\bigl],\\
c_{17} & =8\gEA^{3}\sER^{2}(\left\langle \sER\right\rangle _{1}^{1})^{2},\\
c_{18} & =4\fEA^{3}\fK_{2}\gEA\gEB\sER^{3}(\left\langle \sER\right\rangle _{2}^{0}\left\langle \sER\right\rangle _{1}^{2}-4\left\langle \sER\right\rangle _{1}^{1}\left\langle \sER\right\rangle _{2}^{1}),\\
c_{19} & =4\fEA^{2}\gEA\gEB\sER^{2}\Bigl(\fK_{2}\gEA\sER\bigl[-2(\left\langle \sER\right\rangle _{1}^{1})^{2}-\sER\left\langle \sER\right\rangle _{2}^{0}\left\langle \sER\right\rangle _{1}^{2}+\left\langle \sER\right\rangle _{1}^{1}(2\sER\left\langle \sER\right\rangle _{2}^{1}-\left\langle \sER\right\rangle _{1}^{2})\bigl]\\
 & \quad+2\fEA\gK_{2}\Bigl\{\left\langle \sER\right\rangle _{2}^{1}(2\left\langle \sER\right\rangle _{1}^{1}+\left\langle \sER\right\rangle _{1}^{2})+\sER\bigl[-2(\left\langle \sER\right\rangle _{2}^{1})^{2}+\left\langle \sER\right\rangle _{3}^{0}\left\langle \sER\right\rangle _{1}^{2}\bigl]\Bigl\}\Bigr),\\
c_{20} & =4\fEA^{2}\gEA\sER^{2}(4\left\langle \sER\right\rangle _{1}^{1}\left\langle \sER\right\rangle _{2}^{1}-\left\langle \sER\right\rangle _{2}^{0}\left\langle \sER\right\rangle _{1}^{2}),\\
c_{21} & =-4\fEA\gEA^{2}\sER^{2}\bigl[2(\left\langle \sER\right\rangle _{1}^{1})^{2}+\sER\left\langle \sER\right\rangle _{2}^{0}\left\langle \sER\right\rangle _{1}^{2}+\left\langle \sER\right\rangle _{1}^{1}(\left\langle \sER\right\rangle _{1}^{2}-2\sER\left\langle \sER\right\rangle _{2}^{1})\bigl],\\
c_{22} & =2\fEA^{2}\gEA\sER^{2}\bigl[\left\langle \sER\right\rangle _{1}^{2}(2\sER\left\langle \sER\right\rangle _{2}^{0}-\left\langle \sER\right\rangle _{1}^{2})+4\left\langle \sER\right\rangle _{1}^{1}(\left\langle \sER\right\rangle _{1}^{2}-\sER\left\langle \sER\right\rangle _{2}^{1})\bigl],\\
c_{23} & =2\fEA^{3}\sER^{2}\left\langle \sER\right\rangle _{2}^{1}(\left\langle \sER\right\rangle _{1}^{2}-4\left\langle \sER\right\rangle _{1}^{1}),\\
c_{24} & =2\fEA^{3}\sER^{2}\bigl[4\sER(\left\langle \sER\right\rangle _{2}^{1})^{2}-2\sER\left\langle \sER\right\rangle _{3}^{0}\left\langle \sER\right\rangle _{1}^{2}-3\left\langle \sER\right\rangle _{2}^{1}\left\langle \sER\right\rangle _{1}^{2}\bigl],
\end{align*}
\clearpage~\vspace{-3ex}

\begin{align*}
d_{1} & =2\fEA^{2}\gEA^{3}\sER^{2}\left\langle \sER\right\rangle _{2}^{1}\bigl[\gEB^{2}\gK_{2}^{2}(4\left\langle \sER\right\rangle _{1}^{1}-\left\langle \sER\right\rangle _{1}^{2})-\left\langle \sER\right\rangle _{1}^{2}\bigl],\\
d_{2} & =2\fEA^{2}\gEA^{2}\bigg[2\fK_{2}\gEA\gEB^{2}\gK_{2}\sER^{3}(-4\left\langle \sER\right\rangle _{1}^{1}\left\langle \sER\right\rangle _{2}^{1}+\left\langle \sER\right\rangle _{2}^{0}\left\langle \sER\right\rangle _{1}^{2})+\fEA\Bigl(\sER\left\langle \sER\right\rangle _{2}^{1}\left\langle \sER\right\rangle _{1}^{2}-\left\langle \sER\right\rangle _{1}^{1}\left\langle \sER\right\rangle _{1}^{2}\\
 & \quad+\fK_{2}\gEB^{2}\sER^{3}\Bigl\{-4\fK_{2}\left\langle \sER\right\rangle _{1}^{1}\left\langle \sER\right\rangle _{2}^{1}+\bigl[2\fK_{2}\left\langle \sER\right\rangle _{2}^{0}+(2\fK_{1}+\fK_{2})\left\langle \sER\right\rangle _{2}^{1}\bigl]\left\langle \sER\right\rangle _{1}^{2}\Bigl\}\\
 & \quad\quad+\gEB^{2}\sER^{2}\Bigl\{4\gK_{2}^{2}(\left\langle \sER\right\rangle _{2}^{1})^{2}+\fK_{2}\left\langle \sER\right\rangle _{1}^{2}\bigl[(2\fK_{1}-3\fK_{2})\left\langle \sER\right\rangle _{1}^{1}+2(-\fK_{1}+\fK_{2})\left\langle \sER\right\rangle _{1}^{2}\bigl]\Bigl\}\Bigr)\bigg],\\
d_{3} & =2\fEA^{2}\gEA^{2}\sER\Bigl(\gEA\Bigl\{-2\fK_{2}^{2}\gEB^{2}\sER^{3}\left\langle \sER\right\rangle _{2}^{0}\left\langle \sER\right\rangle _{1}^{2}+\left\langle \sER\right\rangle _{1}^{1}\bigl[4\fK_{2}^{2}\gEB^{2}\sER^{3}\left\langle \sER\right\rangle _{2}^{1}+(1+\fK_{2}^{2}\gEB^{2}\sER^{2})\left\langle \sER\right\rangle _{1}^{2}\bigl]\Bigl\}\\
 & \quad+2\fEA\fK_{2}\gEB^{2}\gK_{2}\sER\Bigl\{2\left\langle \sER\right\rangle _{2}^{1}(\left\langle \sER\right\rangle _{1}^{1}-\left\langle \sER\right\rangle _{1}^{2})+\sER\bigl[-2(\left\langle \sER\right\rangle _{2}^{1})^{2}+\left\langle \sER\right\rangle _{3}^{0}\left\langle \sER\right\rangle _{1}^{2}\bigl]\Bigl\}\Bigr),\\
d_{4} & =-2\fEA^{2}\gEA^{3}\sER\left\langle \sER\right\rangle _{1}^{1}\bigl[\fK_{2}^{2}\gEB^{2}\sER^{2}(4\left\langle \sER\right\rangle _{1}^{1}-3\left\langle \sER\right\rangle _{1}^{2})+\left\langle \sER\right\rangle _{1}^{2}\bigl],\\
d_{5} & =\fEA^{2}\gEA^{3}\gEB^{2}(\left\langle \sER\right\rangle _{1}^{2})^{2}(-2\left\langle \sER\right\rangle _{1}^{1}+2\sER\left\langle \sER\right\rangle _{2}^{1}+\left\langle \sER\right\rangle _{1}^{2}),\\
d_{6} & =-2\fEA^{3}\gEA^{2}\gEB^{2}\left\langle \sER\right\rangle _{1}^{2}\Bigl\{\sER\bigl[(\fJotam 1+2\fJotam 2-2\frhom)\sER\left\langle \sER\right\rangle _{1}^{2}-\left\langle \sER\right\rangle _{3}^{1}\left\langle \sER\right\rangle _{1}^{2}+\left\langle \sER\right\rangle _{2}^{1}(\frhom\sER^{2}-\left\langle \sER\right\rangle _{2}^{2})\bigl]\\
 & \quad+\left\langle \sER\right\rangle _{1}^{1}\bigl[(-2\fJotam 2+\frhom)\sER^{2}+\left\langle \sER\right\rangle _{2}^{2}\bigl]\Bigl\},\\
d_{7} & =2\fEA^{2}\gEA^{3}\gEB^{2}\left\langle \sER\right\rangle _{1}^{2}\Bigl((\sER\left\langle \sER\right\rangle _{2}^{1}-\left\langle \sER\right\rangle _{1}^{2})\left\langle \sER\right\rangle _{1}^{2}+\left\langle \sER\right\rangle _{1}^{1}\Bigl\{\left\langle \sER\right\rangle _{1}^{2}+\sER\bigl[(-2\fJotam 2+\frhom)\sER^{2}+\left\langle \sER\right\rangle _{2}^{2}\bigl]\Bigl\}\Bigr),\\
d_{8} & =4\fEA\fjotam\gEA^{3}\gEB^{2}\sER^{3}\left\langle \sER\right\rangle _{1}^{1}\left\langle \sER\right\rangle _{1}^{2},\\
d_{9} & =2\fEA^{2}\gEA^{3}\gEB^{2}\sER\left\langle \sER\right\rangle _{1}^{1}\left\langle \sER\right\rangle _{1}^{2}(\frhom\sER^{2}-\left\langle \sER\right\rangle _{2}^{2}),\\
d_{10} & =\fEA^{2}\gEA^{3}\gEB^{2}\sER^{2}\left\langle \sER\right\rangle _{1}^{2}\bigl[2\left\langle \sER\right\rangle _{2}^{1}(\grhom-\left\langle \sER\right\rangle _{0}^{2})-4\left\langle \sER\right\rangle _{1}^{1}\left\langle \sER\right\rangle _{1}^{2}+(\left\langle \sER\right\rangle _{1}^{2})^{2}\bigl],\\
d_{11} & =-6\fEA^{3}\gEA^{2}\gEB^{2}\sER^{2}\left\langle \sER\right\rangle _{2}^{1}(\left\langle \sER\right\rangle _{1}^{2})^{2},\\
d_{12} & =4\fEA\gEA^{3}\gEB\gK_{2}\sER^{2}(-4\left\langle \sER\right\rangle _{1}^{1}\left\langle \sER\right\rangle _{2}^{1}+\left\langle \sER\right\rangle _{2}^{0}\left\langle \sER\right\rangle _{1}^{2}),\\
d_{13} & =4\fEA\gEA^{2}\gEB\sER^{2}\Bigl\{\fEA\gK_{2}\bigl[-4(\left\langle \sER\right\rangle _{2}^{1})^{2}+\left\langle \sER\right\rangle _{3}^{0}\left\langle \sER\right\rangle _{1}^{2}\bigl]\\
 & \quad-2\fK_{2}\gEA\bigl[(\left\langle \sER\right\rangle _{1}^{1})^{2}+\sER\left\langle \sER\right\rangle _{2}^{0}\left\langle \sER\right\rangle _{1}^{2}-\left\langle \sER\right\rangle _{1}^{1}(3\sER\left\langle \sER\right\rangle _{2}^{1}+\left\langle \sER\right\rangle _{1}^{2})\bigl]\Bigl\},\\
d_{14} & =2\fEA\gEA^{2}\Bigl\{\left\langle \sER\right\rangle _{1}^{1}\left\langle \sER\right\rangle _{1}^{2}-\sER\bigl[4\sER(\left\langle \sER\right\rangle _{2}^{1})^{2}-2\sER\left\langle \sER\right\rangle _{3}^{0}\left\langle \sER\right\rangle _{1}^{2}+\left\langle \sER\right\rangle _{2}^{1}\left\langle \sER\right\rangle _{1}^{2}\bigl]\Bigl\},\\
d_{15} & =-2\gEA^{3}\sER\left\langle \sER\right\rangle _{1}^{1}(-4\sER\left\langle \sER\right\rangle _{2}^{1}+\left\langle \sER\right\rangle _{1}^{2}),\\
d_{16} & =2\gEA^{3}\sER\bigl[-2\sER\left\langle \sER\right\rangle _{2}^{0}\left\langle \sER\right\rangle _{1}^{2}+\left\langle \sER\right\rangle _{1}^{1}(4\sER\left\langle \sER\right\rangle _{2}^{1}+\left\langle \sER\right\rangle _{1}^{2})\bigl],\\
d_{17} & =-8\fEA^{3}\fK_{2}\gEA\gEB\sER^{2}\left\langle \sER\right\rangle _{2}^{1}\left\langle \sER\right\rangle _{1}^{2},\\
d_{18} & =4\fEA^{2}\fK_{2}\gEA^{2}\gEB\sER^{3}(-4\left\langle \sER\right\rangle _{1}^{1}\left\langle \sER\right\rangle _{2}^{1}+\left\langle \sER\right\rangle _{2}^{0}\left\langle \sER\right\rangle _{1}^{2}),\\
d_{19} & =4\fEA^{3}\fK_{2}\gEA\gEB\sER^{2}\Bigl\{-2\left\langle \sER\right\rangle _{2}^{1}(\left\langle \sER\right\rangle _{1}^{1}-2\left\langle \sER\right\rangle _{1}^{2})+\sER\bigl[2(\left\langle \sER\right\rangle _{2}^{1})^{2}-\left\langle \sER\right\rangle _{3}^{0}\left\langle \sER\right\rangle _{1}^{2}\bigl]\Bigl\},\\
d_{20} & =-8\fEA^{3}\fK_{2}\gEA\gEB\sER^{2}\left\langle \sER\right\rangle _{2}^{1}\left\langle \sER\right\rangle _{1}^{2},\\
d_{21} & =-4\fEA\gEA^{2}\sER^{2}\Bigl\{\left\langle \sER\right\rangle _{2}^{1}(2\left\langle \sER\right\rangle _{1}^{1}+\left\langle \sER\right\rangle _{1}^{2})+\sER\bigl[-2(\left\langle \sER\right\rangle _{2}^{1})^{2}+\left\langle \sER\right\rangle _{3}^{0}\left\langle \sER\right\rangle _{1}^{2}\bigl]\Bigl\},\\
d_{22} & =4\fEA^{2}\gEA\sER^{2}\bigl[4(\left\langle \sER\right\rangle _{2}^{1})^{2}-\left\langle \sER\right\rangle _{3}^{0}\left\langle \sER\right\rangle _{1}^{2}\bigl],\\
d_{23} & =2\fEA^{2}\gEA\sER^{2}\bigl[-4\sER(\left\langle \sER\right\rangle _{2}^{1})^{2}+2\sER\left\langle \sER\right\rangle _{3}^{0}\left\langle \sER\right\rangle _{1}^{2}+3\left\langle \sER\right\rangle _{2}^{1}\left\langle \sER\right\rangle _{1}^{2}\bigl],\\
d_{24} & =-8\fEA^{3}\sER^{2}(\left\langle \sER\right\rangle _{2}^{1})^{2},
\end{align*}
\clearpage

\section{On solving the momentum constraint for $p$}

\label{app:dtp}

In spherical symmetry, we can solve either of the momentum constraints
$\gSector{\mathcal{C}}_{i}$ or $\fSector{\mathcal{C}}_{i}$ for $\sLp$.
In general, one can solve one of these constraints for $\sLp$ numerically
at every time step of the numerical integration. As we described in
the main text, it would be desirable to have, if not an exact expression
for $\sLp$, at least an exact expression for $\partial_{t}\sLp$,
since this would allow us to evolve in time \emph{any} object entering
in the bimetric decomposition introduced in \cite{Kocic:2018ddp}.
Indeed, $\sLp$ enters in all the definitions of the 3+1 bimetric
interactions.

In this appendix we review one strategy used to try to compute $\sLp$.
Unfortunately, it did not work, but we think it is good to show this
method both to prevent other interested people from trying it and
to give some hints about what can be done.

The approach concerns solving the momentum constraints for $\sLp$
in full generality. Let's consider the momentum constraint in the
$g$-sector \eqref{eq:g-cc-vector},
\begin{align}
\gCC_{i} & =\gD_{k}\gK^{k}{}_{i}-\gD_{i}\gK\signK\gKappa\rb{\gjota_{i}^{\gblab}+\gjota_{i}^{\gmlab}}=0\quad\Longrightarrow\quad\gjota_{i}^{\gblab}=\gKappa^{-1}\rb{\gD_{k}\gK^{k}{}_{i}-\gD_{i}\gK}-\gjota_{i}^{\gmlab}.
\end{align}
Suppose that we can write the bimetric current as $\gjota_{i}^{\gblab}=\sLp_{a}\tud{\mathbb{M}}ai$,
with $\tud{\mathbb{M}}ai$ some linear operator independent of $\sLp$.
Then, if $\mathbb{M}$ is invertible, we have an exact expression
for $\sLp$ in terms of the other fields, 
\begin{equation}
\sLp_{a}=\tud{(\mathbb{M}^{-1})}ia\qb{\gKappa^{-1}\rb{\gD_{k}\gK^{k}{}_{i}-\gD_{i}\gK}-\gjota_{i}^{\gmlab}}\label{eq:p}
\end{equation}
We have not been able to find such a matrix $\mathbb{M}$ so far,
and the following calculations describe our approach.

The bimetric current $\gjota_{i}^{\gblab}$ is given by 
\begin{equation}
\gjota_{i}^{\gblab}=\usignV\gSp_{ij}\sgQU^{j}{}_{k}\sgn^{k}=-\gSp_{ij}\sgQU^{j}{}_{k}(\gE^{-1})^{k}{}_{a}\,\sLp^{a}\sLt^{-1}.\label{eq:bimcurrent}
\end{equation}
Note that $\sLp$ is inside $\sgQU^{j}{}_{k}$ and therefore we cannot
just invert  $\gSp_{ij}\sgQU^{j}{}_{k}(\gE^{-1})^{k}{}_{a}$ to find
$\sLp$. We need to extract it from $\sgQU^{j}{}_{k}$ (if possible).
We start by explicitly writing down the quantity $\gSp_{ij}\sgQU^{j}{}_{k}$
in matrix notation, 
\begin{align}
\gSp\sgQU & =\usignV\betaSum\,\gSp\sgB\,Y_{n-1}(\sfD)\nonumber \\
 & =\usignV m^{4}\rb{\beta_{(0)}\hSp\,Y_{-1}(\sfD)+\beta_{(1)}\hSp\,Y_{0}(\sfD)+\beta_{(2)}\hSp\,Y_{1}(\sfD)+\beta_{(3)}\hSp\,Y_{2}(\sfD)+\beta_{(4)}\sgB\,Y_{3}(\sfD)},
\end{align}
where $Y_{n}(\sfD)\coloneqq\sum\limits _{k=0}^{n}(-1)^{n+k}e_{k}(\sfD)\sfD^{n-k}$,
$Y_{-1}(\sfD)=0$ by definition and $Y_{3}(\sfD)=0$ due to the Cayley\textendash Hamilton
theorem. The other terms read,\bSe 
\begin{align}
\hSp Y_{0}(\sfD) & =\hSp,\\
\hSp Y_{1}(\sfD) & =-\fSp+e_{1}(\sfD)\hSp,\\
\hSp Y_{2}(\sfD) & =-\fSp\sfD-e_{1}(\sfD)\fSp+e_{2}(\sfD)\hSp,
\end{align}
\eSe where $\hSp=\gE^{\intercal}\sEta\sLs\fE=\hSp^{\intercal}$,
which contains $\sLp$. 

Starting with $\hSp Y_{0}(\sfD)$, we can write, 
\begin{align}
\hSp Y_{0}(\sfD)\gE^{-1}\sLp\sLt^{-1} & =\hSp\gE^{-1}\sLp\sLt^{-1}=\hSp^{\intercal}\gE^{-1}\sLp\sLt^{-1}=\fE^{\intercal}\sLs^{\intercal}\sEta\gE\gE^{-1}\sLp\sLt^{-1}\nonumber \\
 & =\fE^{\intercal}\sEta\sLp\sLt^{-1}+\dfrac{1}{\sLt(\sLt+1)}\sEta\sLp\rb{\sLp^{\intercal}\sEta\sLp}=\dfrac{1}{\sLt}\rb{\fE^{\intercal}+(\sLt-1)\sI}\sEta\sLp.\label{eq:Y0}
\end{align}
 Note that $\sLt=\sqrt{1+\sLp^{\intercal}\sEta\sLp}$, hence this
expression is not very convenient if we want to solve for $\sLp$.
An idea would be to treat $\sLt$ and $\sLp$ as two independent variables
and solve the \emph{two} momentum constraints, one for $\sLt$ and
the other for $\sLp$, if possible.

The second term can be rewritten as 
\begin{align}
\hSp Y_{1}(\sfD)\gE^{-1}\sLp\sLt^{-1} & =-\fSp\gE^{-1}\sLp\sLt^{-1}+e_{1}(\sfD)\rb{\hSp\gE^{-1}\sLp\sLt^{-1}}\nonumber \\
 & =-\fSp\gE^{-1}\sLp\sLt^{-1}+e_{1}(\sfD)\qb{\dfrac{1}{\sLt}\rb{\fE^{\intercal}+(\sLt-1)\sI}\sEta\sLp}\nonumber \\
 & =\qb{-\fSp\gE^{-1}+e_{1}(\sfD)\rb{\fE^{\intercal}+(\sLt-1)\sI}\sEta}\sLp\sLt^{-1},\label{eq:Y1}
\end{align}
where we used \eqref{eq:Y0} to write the second equality. We were
not able to isolate $\sLp$ more than this. Note that $\sLp$ is also
inside $e_{1}(\sfD)=\Tr(\sfD)$. We can rewrite this trace as, 
\begin{align}
\Tr(\sfD) & =\Tr(\gE^{-1}\sLs^{-1}\fE)=\Tr\qb{\gE^{-1}\rb{\sI-\dfrac{1}{\sLt(\sLt+1)}\sLp\sLp^{\intercal}\sEta}\fE}\nonumber \\
 & =\Tr\qb{\gE^{-1}\fE}-\dfrac{1}{\sLt(\sLt+1)}\Tr\qb{\gE^{-1}\sLp\sLp^{\intercal}\sEta\fE}.\label{eq:e2}
\end{align}
The cyclic property of the trace tells us that 
\begin{align}
\Tr\qb{\gE^{-1}\sLp\sI\sLp^{\intercal}\sEta\fE}=\Tr\qb{\sLp^{\intercal}\sEta\fE\gE^{-1}\sLp\sI}=\sLp^{\intercal}\sEta\fE\gE^{-1}\sLp\Tr\qb{\sI}=3\,\sLp^{\intercal}\sEta\fE\gE^{-1}\sLp.\label{eq:e1}
\end{align}
The substitution of \eqref{eq:e2} and \eqref{eq:e1} in \eqref{eq:Y1}
gives 
\begin{align}
\hSp Y_{1}(\sfD)\gE^{-1}\sLp\sLt^{-1} & =\qb{-\fSp\gE^{-1}+e_{1}(\sfD)\rb{\fE^{\intercal}+(\sLt-1)\sI}\sEta}\sLp\sLt^{-1}\nonumber \\
 & =\qb{-\fSp\gE^{-1}+\rb{\Tr\qb{\gE^{-1}\fE}-\dfrac{3\,\sLp^{\intercal}\sEta\fE\gE^{-1}\sLp}{\sLt(\sLt+1)}}\rb{\fE^{\intercal}+(\sLt-1)\sI}\sEta}\sLp\sLt^{-1}.\label{eq:Y1final}
\end{align}

The third term is, 
\begin{align}
\hSp Y_{2}(\sfD)\gE^{-1}\sLp\sLt^{-1} & =-\fSp\sfD\gE^{-1}\sLp\sLt^{-1}-e_{1}(\sfD)\fSp\gE^{-1}\sLp\sLt^{-1}+e_{2}(\sfD)\rb{\hSp\gE^{-1}\sLp\sLt^{-1}}\nonumber \\
 & =-\fSp\sfD\gE^{-1}\sLp\sLt^{-1}-e_{1}(\sfD)\fSp\gE^{-1}\sLp\sLt^{-1}+e_{2}(\sfD)\qb{\dfrac{1}{\sLt}\rb{\fE^{\intercal}+(\sLt-1)\sI}\sEta\sLp}\nonumber \\
 & =-\fSp\gE^{-1}\rb{\sI-\dfrac{1}{\sLt(\sLt+1)}\rb{\sLp^{\intercal}\sEta\fE\gE^{-1}\sLp}\gE\fE^{-1}}\fE\gE^{-1}\sLp\sLt^{-1}\nonumber \\
 & \quad-e_{1}(\sfD)\fSp\gE^{-1}\sLp\sLt^{-1}+e_{2}(\sfD)\qb{\dfrac{1}{\sLt}\rb{\fE^{\intercal}+(\sLt-1)\sI}\sEta\sLp},\label{eq:Y2}
\end{align}
where the last equality follows after doing some straightforward algebra.
We need to compute $e_{2}(\sfD)=\dfrac{1}{2}\qb{\Tr(\sfD)^{2}-\Tr(\sfD^{2})}$.
We already know the expression for $\Tr(\sfD)$, hence we turn our
attention to $\Tr(\sfD^{2})$. First, we rewrite $\sfD^{2}$ in a
more handy way, 
\begin{align}
\sfD=\sgD\gSp^{-1}\fSp=\gSp^{-1}\sgD^{\intercal}\fSp\Longrightarrow\sfD^{2} & =\gSp^{-1}\sgD^{\intercal}\fSp\sfD=\gSp^{-1}\sgD^{\intercal}\sfD^{\intercal}\fSp\nonumber \\
 & =\gSp^{-1}\gE^{\intercal}\sLs^{-1,\intercal}\fE^{-1,\intercal}\fE^{\intercal}\sLs^{-1,\intercal}\gE^{-1,\intercal}\fSp\\
 & =\gE^{-1}\sEta^{-1}\gE^{-1,\intercal}\gE^{\intercal}\sLs^{-2,\intercal}\gE^{-1,\intercal}\fSp\\
 & =\gE^{-1}\sEta^{-1}\sLs^{-2,\intercal}\gE^{-1,\intercal}\fSp.
\end{align}
Now $\sLs$ only appears once inside $\sfD^{2}$. Recalling that $\sLs^{-2}=\sI-\sLt^{-2}\sLp\sLp^{\intercal}\sEta$,
we can compute the trace of $\sfD^{2}$, 
\begin{align}
\Tr(\sfD^{2}) & =\Tr\qb{\gE^{-1}\sEta^{-1}\sLs^{-2,\intercal}\gE^{-1,\intercal}\fSp}=\Tr\qb{\gSp^{-1}\fSp}-\sLt^{-2}\Tr\qb{\gE^{-1}\sLp\sLp^{\intercal}\gE^{-1,\intercal}\fSp}.
\end{align}
The cyclic property of the trace tells us that, 
\begin{equation}
\Tr\qb{\gE^{-1}\sLp\sI\sLp^{\intercal}\gE^{-1,\intercal}\fSp}=\Tr\qb{\sLp^{\intercal}\gE^{-1,\intercal}\fSp\gE^{-1}\sLp\sI}=\sLp^{\intercal}\gE^{-1,\intercal}\fSp\gE^{-1}\sLp\Tr\qb{\sI}=3\,\sLp^{\intercal}\gE^{-1,\intercal}\fSp\gE^{-1}\sLp.\label{eq:trD2}
\end{equation}
The substitution of \eqref{eq:trD2} in \eqref{eq:Y2} becomes 
\begin{align}
\hSp Y_{2}(\sfD)\gE^{-1}\sLp\sLt^{-1} & =-\fSp\gE^{-1}\rb{\sI-\dfrac{1}{\sLt(\sLt+1)}\rb{\sLp^{\intercal}\sEta\fE\gE^{-1}\sLp}\gE\fE^{-1}}\fE\gE^{-1}\sLp\sLt^{-1}\nonumber \\
 & \quad-\rb{\Tr\qb{\gE^{-1}\fE}-\dfrac{3\,\sLp^{\intercal}\sEta\fE\gE^{-1}\sLp}{\sLt(\sLt+1)}}\fSp\gE^{-1}\sLp\sLt^{-1}\nonumber \\
 & \quad+\dfrac{1}{2}\left\lbrace \vphantom{\dfrac{\sLp^{\intercal}\sEta\fE\gE^{-1}\sLp}{\sLt(\sLt+1)}}\Tr\rb{\fE\gE^{-1}}^{2}-\Tr\qb{\rb{\fE\gE^{-1}}'\fE\gE^{-1}}+3\sLt^{-2}\sLp^{\intercal}\gE^{-1,\intercal}\fSp\gE^{-1}\sLp\right.\nonumber \\
 & \qquad\quad\;\,+\left.\dfrac{\sLp^{\intercal}\sEta\fE\gE^{-1}\sLp}{\sLt(\sLt+1)}\qb{\dfrac{\sLp^{\intercal}\sEta\fE\gE^{-1}\sLp}{\sLt(\sLt+1)}-2\Tr\rb{\fE\gE^{-1}}}\right\rbrace \cdot\nonumber \\
 & \quad\cdot\qb{\dfrac{1}{\sLt}\rb{\fE^{\intercal}+(\sLt-1)\sI}\sEta\sLp}.\label{eq:Y2final}
\end{align}

Now we know the explicit expressions for all the terms contributing
to $\gjotab$ in \eqref{eq:bimcurrent}, as functions of $\sLp$ and
$\sLt=\sqrt{1+\sLp^{\intercal}\sEta\sLp}$. We have, for the covector
$\gjotab$, 
\begin{align}
\gjotab & =-\gSp\sgQU\sgn\nonumber \\
 & =m^{4}\qb{\beta_{(1)}\rb{\hSp Y_{0}(\sfD)\gE^{-1}\sLp\sLt^{-1}}+\beta_{(2)}\rb{\hSp Y_{1}(\sfD)\gE^{-1}\sLp\sLt^{-1}}+\beta_{(3)}\rb{\hSp Y_{2}(\sfD)\gE^{-1}\sLp\sLt^{-1}}},
\end{align}
where the terms in the round brackets are given by \eqref{eq:Y0},
\eqref{eq:Y1final}, \eqref{eq:Y2final}, respectively. In the end,
one cannot state if there exists a matrix $\mathbb{M}$ satisfying
\eqref{eq:p}. We hope that this method can be helpful for further
attempts to solve this problem.

%%%%%%%%%%%%%%%%%%%%%%%%%%%%%%%%%%%%%%%%%%%%%%%%%%%%%%%%%%%%%%%%%%%%%%%%%%%%%
% Bibliography
%%%%%%%%%%%%%%%%%%%%%%%%%%%%%%%%%%%%%%%%%%%%%%%%%%%%%%%%%%%%%%%%%%%%%%%%%%%%%

\clearpage
\ifprstyle

\bibliographystyle{apsrev4-1}
\bibliography{lapse-ratio}

\else

\bibliographystyle{JHEP}
\bibliography{lapse-ratio}

\fi

\end{document}